\documentclass[11pt,a4paper]{article}
\pdfoutput=1 
\usepackage{jheppub}
\usepackage[T1]{fontenc} 

\usepackage{amsmath}
\usepackage{amssymb}
\usepackage{mathrsfs}
\usepackage{graphicx}

\usepackage{enumerate}

\def\half{\frac{1}{2}}
\def\eq#1 { \begin{equation} #1 \end{equation} }
\def\eqn#1{ \begin{eqnarray} #1 \end{eqnarray} }
\def\nn { \nonumber }
\def\Reals{\mathbb{R}}

\def\Re{{\rm Re}\,}
\def\Im{{\rm Im}\,}
\def\a{\alpha}

\def\d{\partial}
\def\cH{\mathcal{H}}
\def\cN{\mathcal{N}}
\def\cA{\mathcal{A}}
\def\cD{\mathcal{D}}
\def\cO{\mathcal{O}}
\def\cM{\mathcal{M}}
\def\cB{\mathcal{B}}
\def\cS{\mathcal{S}}

\def\vol{{\rm Vol}}

\def\C#1{\left\langle #1 \right\rangle}
\def\MI{{\rm MI}}
\def\TMI{{\rm TMI}}
\def\HMI{{\rm HMI}}
\def\HTMI{{\rm HTMI}}

\newcommand{\tr}{{\mathrm{tr}}}
 \newcommand{\be}{\begin{equation}}
\newcommand{\ee}{\end{equation}}

\newcommand{\ket}[1]{\left| #1 \right>} 
\newcommand{\bra}[1]{\left< #1 \right|} 

\newcommand{\shr}[2]{\sinh\left[\frac{#1}{#2}\right]}

\title{\vspace{-1in}Mutual information between thermo-field doubles
  and disconnected holographic boundaries}

\author[a]{Ian A. Morrison}
\author[b]{and Matthew M. Roberts}

\affiliation[a]{DAMTP, Centre of Mathematical Sciences,
     University of Cambridge, U.K.}
\affiliation[b]{Department of Physics and Center for Cosmology 
     and Particle Physics, New York University, New York, New York, USA}

\emailAdd{i.morrison@damtp.cam.ac.uk}
\emailAdd{matthew.roberts@nyu.edu}

\abstract{
  We use mutual information as a measure of the entanglement between `physical' and thermo-field double degrees of freedom in field theories at finite temperature. We compute this ``thermo-mutual information'' in simple toy models: a quantum mechanics two-site spin chain, a two dimensional massless fermion, and a two dimensional holographic system. In holographic systems, the thermo-mutual information is related to minimal surfaces connecting the two disconnected boundaries of an eternal black hole.
  We derive a number of salient features of this thermo-mutual information, including that it is UV finite, positive definite and bounded from above by the standard mutual information for the thermal ensemble. 
  We relate the construction of the reduced density matrices used to define the thermo-mutual information to the Schwinger-Keldysh formalism,  ensuring that all our objects are well defined in Euclidean and Lorentzian signature. 
}

\begin{document}
\maketitle





\section{Introduction}
\label{sec:intro}

Mutual information (MI) provides a valuable measure of the entanglement 
of a quantum system. Unlike the von Neumann entanglement entropy of 
spatial regions, mutual information does not generically suffer UV 
divergences in quantum field theory (QFT). 
The mutual information between two subregions $A$ and $B$ describes
the total correlation, both classical and quantum, between the regions
\cite{Groisman:2005aa}; in particular, the mutual information provides 
an upper bound on the correlation for any two observables 
supported in $A$ and $B$ respectively \cite{Wolf:2008aa}. 
As a result the mutual information may be used to ascertain basic
properties of any local observable in a quantum system.

Unfortunately, a practical method of computing MI for arbitrary regions in a generic QFT does not exist, making detailed analysis of MI in systems of interest quite difficult. Exact analytic computations of MI are rare and often rely heavily on conformal symmetry (see e.g. \cite{Calabrese:2004eu,Calabrese:2009qy,Casini:2009sr} and references therein). Following the conjecture of Ryu and Takayanagi \cite{Ryu:2006bv} it is widely believed that the AdS/CFT correspondence can be used to compute quantum informatics quantities in certain strongly 
coupled QFTs. 
In particular, it is believed that the Ryu-Takayanagi (RT) formula 
\cite{Ryu:2006bv} for the holographic entanglement entropy  captures the 
leading $N^2$ contribution the entanglement entropy of the dual CFT.
Efforts have been made towards a concrete derivation of the RT formula
(see, e.g., \cite{Ryu:2006bv,Michalogiorgakis:2008aa,Schwimmer:2008aa,
Headrick:2010zt,Casini:2011kv}) but this formula and its covariant 
formulation \cite{Hubeny:2007xt} remain a conjecture. 
Furthermore, currently there does not exist a tractable 
method for calculating subleading corrections to the RT formula
(see for instance \cite{Headrick:2010zt}). However, the robustness of the 
RT formula has inspired a great deal of investigation
into the relationship between quantum information and holography
\cite{Hubeny:2012ry,Hubeny:2012wa,Czech:2012bh,Bousso:2012sj}.

In this paper we examine the holographic mutual information between regions on disconnected boundaries in the maximally extended eternal Schwarzschild-AdS spacetime. The standard holographic interpretation of asymptotically AdS spacetimes with multiple disconnected boundaries is that they provide a bulk realization of an entangled state in a system of multiple non-interacting field theories
\footnote{The robustness of this interpretation is debated \cite{Marolf:2012xe}.}. In the simplest case of a maximally extended Schwarzschild-AdS black hole, which has two causally disconnected boundaries, one can interpret one boundary as the thermo-field double of the field theory on the second boundary \cite{Maldacena:2001kr}. Therefore, the mutual information we study holographically should characterize the mutual information between operators and their thermo-field double -- we call this type of mutual information ``thermo-mutual information'' so as to distinguish it from mutual information for purely `physical' operators. The thermo-mutual information quantity provides information about the level of entanglement of the thermal system. It is our hope that this work will lay necessary foundation for a more broad study of holographic mutual information between regions on disconnected boundaries of bulk spacetimes.  

As a warm-up to our analysis of holographic systems, we compute the mutual information between `physical' fields and their thermo-field doubles in two simple systems: a quantum mechanical two-site spin chain and a  two-dimensional massless fermion. We then calculate thermo-mutual information for a 2d holographic system dual to asymptotically $AdS_3$ spacetime in the simple case of a non-rotating BTZ black hole. We find similar behavior in these three very different systems.  

The rest of the paper is organized as follows. Section~\ref{sec:2} is devoted to introducing 
thermo-mutual information in the setting of quantum field theory 
without regard to holographic duals.
We begin by briefly reviewing mutual information and thermal
quantum systems in \S\ref{sec:review} and \S\ref{sec:thermal} respectively;
we then define thermo-mutual information and describe its basic 
attributes in \S\ref{sec:TMI}. We provide two example
calculations of TMI: the first in the comfortable setting of a
quantum mechanical spin chain (\S\ref{sec:TMI}), and the
second, after describing some computational details in
\S\ref{sec:QFT} (some derivations are in Appendix \ref{app:PI} for brevity), in a 2D CFT (\S\ref{sec:CFT}). Section~\ref{sec:HTMI} is devoted to studying holographic systems. We first review the conjecture of Ryu and Takayanagi in, and give a brief summary of the non rotating BTZ geometry, laying out the tools which make our calculation  straightforward. The explicit calculation of mutual information and thermo-mutual information is in \S\ref{sec:computation}. We close with a discussion and comments on generalizations of our calculations in section \ref{sec:discussion}. 

\section{Thermo-mutual information in field theory}
\label{sec:2}

Unless stated otherwise, throughout this section we consider 
QFTs on a space $\cM = \mathbb{R}\times \Lambda$ 
with $\mathbb{R}$ the continuous time and $\Lambda$ a $(d-1)$-dimensional spatial 
lattice with spacing $a$, which we may simply take to be square.
The QFT has a Hilbert space $\cH$ and an algebra of local observables 
$\cA(\cM)$ represented by bounded operators on $\cH$. 
The reason for considering a lattice theory is that for such theories
the Hilbert space that is explicitly a direct product of the Hilbert space 
at each site, $\cH = \otimes_\Lambda \cH_{site}$, and therefore may be 
decomposed into spatial regions.
The support of a local observable may be as small as a single
lattice site labelled by the position vector $x$.
We expect to be able to address continuum QFTs by taking the $a\to 0$ limit;
more carefully, we could say that we define a continuum QFT by this limit.

\subsection{Mutual information}
\label{sec:review}

Mutual information is constructed from reduced density matrices,
so we begin by discussing these objects.
Quite in general a quantum state $\Psi$ may be described by its 
density matrix $\rho := \rho(\Psi; \cA(\cM))$.
The lengthy notation denotes that $\rho$ is properly defined as 
a functional on $\cA(\cM)$ which defines the state $\Psi$.
Correlation functions of an observable $\cO \in \cA(\cM)$ with 
respect to $\Psi$ may be computed by tracing against $\rho$:
\eq{
  \C{ \cO }_\Psi := \tr_\cH[ \rho \,\cO ] .
}
However, if we restrict our attention to observables supported 
in a subset $A \subset \Sigma_t$ with $\Sigma_t$ an equal-time
slice on $\cM$, then we may compute correlation 
functions without knowing $\rho$ in its entirety;
it suffices to have the reduced density matrix 
$\rho_A := \rho(\Psi; \cA(A))$. 
Due to the underlying lattice structure of $\cM$
the Hilbert space may be factorized in position space, i.e.
$\cH = \cH_A \times \cH_{\overline{A}}$ where $\overline{A}$
is the complement of $A$ on $\Sigma_t$.
Then $\rho_A$ is obtained by tracing over $\cH_{\overline{A}}$:
\eq{
  \rho_{A} = \tr_{\cH_{\overline{A}}}[\rho ].
}
Correlation functions of any observable $\cO_A \in \cA(A)$ 
may then be obtained from operations in $\cH_A$ alone via
\eq{
  \C{ \cO_A }_\Psi = \tr_{\cH_A}[\rho_{A} \cO_A ] .
}

The mutual information for two disjoint subsets 
$A, B \subset \Sigma_t$ is given by
\eq{ \label{eq:I}
  \MI(A : B) := S(\rho_A) + S(\rho_B) - S(\rho_{A \cup B}) \ge 0,
}
where $S(\rho)$ is the entanglement entropy (a.k.a.~Von Neumann entropy) 
associated to a density matrix:
\eq{ \label{eq:S}
  S(\rho) := - \tr[ \rho \log \rho ] .
}
The trace is over the Hilbert space for which $\rho$ is a trace
operator. 
The final inequality in (\ref{eq:I}) is a consequence of the fact that 
entanglement entropy satisfies strong subadditivity, i.e., 
\eq{ \label{eq:SSA}
  S(A) + S(B) \ge S(A\cup B) .
}
Mutual information quantifies the total correlation, both classical
and quantum, between the subsets $A$ and $B$ \cite{Groisman:2005aa}.
This is seen, for instance, in the Pinsker inequality
\cite{Wolf:2008aa}:
\eq{ \label{eq:IPinsker}
  \MI(A : B) \ge \half \left(
  \frac{\C{\cO_A \cO_B}_\Psi - \C{\cO_A}_\Psi \C{\cO_B}_\Psi}
  {||\cO_A|| ||\cO_B||}\right)^2 ,
  \quad \cO_A \in \cA(A), \; \cO_B \in \cA(B) .
}

Mutual information may be defined for continuum QFTs by taking the 
limit $a \to 0$. In this limit the entanglement entropy $S(\rho)$ is 
UV divergent, under reasonable conditions \cite{Casini:2003ix} scaling as $S(\rho_A) \sim\mathrm{Area}(\partial A)/a^{d-2}+\ldots$. For $d=2$ QFTS, which will be what we study in detail, the divergent part of $S(\rho)$ has the universal scaling behavior
\eq{
  S(\rho_A) \sim \gamma c N \log a .
}
Here $\gamma$ is a constant which depends on the theory but not
the specifics of the region $A$, $c$ is number which characterized
the ``number of effective degrees of freedom'' (i.e. the central charge
in a CFT), and $N$ is the number of boundary points of $A$.
As a result of this universal behavior we may construct a UV-finite
symmetric function \cite{Casini:2004bw}:
\eq{ \label{eq:F}
  F(A : B) = S(\rho_{A}) + S(\rho_{A}) - S(\rho_{A\cup B}) 
  - S(\rho_{A \cap B}) .
}
The mutual information agrees with this function within MI's domain
of definition $A\cap B = \emptyset$.
Similar statements a believed to hold generically in higher dimensions as well, as all divergent terms are expected to be expressible as integrals over $\partial A$ that only depend on the cutoff scale and therefore should cancel
-- see discussion in \cite{Calabrese:2009qy,Eisert:2010fk,Refael:2009aa}
--  though these statements are tempered by the fact that there exist
few exact calculations of entanglement entropy.

\subsection{Thermal systems}
\label{sec:thermal}

The canonical description of thermal states in QFT follows from
the semi-classical interpretation of these states as equilibrium states 
in a grand canonical ensemble. Consider a system governed by a 
time-independent Hamiltonian $H$. 
The thermal state $\Omega$ with inverse temperature $\beta$
is defined by a density matrix $\rho = e^{-\beta H}$;
normalized correlations functions of the state are constructed via:
\eq{ \label{eq:Omega1}
  \C{ \phi(x) \ldots}_\Omega := \frac{1}{Z} \tr[\rho \phi(x)\ldots] 
  = \frac{1}{Z} \tr[ e^{-\beta H} \phi(x) \ldots] ,
}
where $Z := \tr[\rho]$ so that $\C{1}_\Omega = 1$.
This description of $\Omega$ provides a simple connection to
classical statistical mechanics.

However, it is often preferable to describe $\Omega$ as a pure rather than 
mixed state. This may be accomplished with the formalism now known as
``thermo-field dynamics'' (TFD) -- original works include
\cite{Schwinger:1960qe,Keldysh:1964ud}, for useful reviews see 
\cite{Landsman:1986uw,Takahashi:1996zn}.
TFD employs an enlarged quantum system composed of two copies of 
the original Hilbert space $\cH$.
The total Hilbert space of this system is given by the tensor product
\eq{
  \cH_{\rm TFD} = \cH_1 \otimes \cH_2,
}
with $\cH_1$, $\cH_2$ isomorphic to $\cH$. The total Hamiltonian
is
\eq{ \label{eq:Htot}
  H_{\rm TFD} = H_1 \otimes 1_2 - 1_1 \otimes H_2 ,
}
and $\Omega$ may be written in the product space as
\eq{ \label{eq:Omega2}
  \ket{\Omega} = \frac{1}{\sqrt{Z}} \sum_n e^{-\beta E_n/2}
  \ket{E_n}_1 \otimes \ket{E_n}_2 ,
}
where $\ket{E_n}_i$ are eigenstates of $H_i$ belonging to
$\cH_i$ respectively.
From (\ref{eq:Htot}) and (\ref{eq:Omega2}) we readily see that
$H_{\rm TFD} \ket{\Omega} = 0$.
For every operator $\phi(x)$ of the original theory there are
two copies $\phi_1(x)$, $\phi_2(x)$ in the `purified' theory.
Correlators involving only type-1 operators reproduce correlators
of the original theory, e.g.,
\eq{
  \bra{\Omega}\phi_1(x)\phi_1(y)\dots\ket{\Omega} 
  = \C{\phi(x)\phi(y)\dots}_\Omega .
}
Since the Hamiltonian does not mix operators on $\cH_1$
and $\cH_2$, all type-1 operators commute with all type-2 operators:
\eq{
  \bra{\Omega}\dots \left[\phi_1(x), \phi_2(y)\right] \dots \ket{\Omega} = 0 .
}

The algebra of observables of the TFD system is
\eq{ \label{eq:ATFD}
  \cA_{\rm TFD}(\cM) := \cA_1(\cM) \otimes \cA_2(\cM) .
}
It is not always conventional to regard type-2 operators as ``observable.'' 
Because correlators of type-1 operators reproduce correlators of the
original theory, the second copy of the field theory is sometimes
referred to as just a `useful fiction' or computational trick.
Furthermore, correlators of exclusively type-2 operators may be related to 
correlators of exclusively type-1 operators via a anti-unitary map.
Ultimately, our motivation for placing the type-2 operators on equal
footing with those of type-1 originates from the dual description in
terms of eternal black hole spacetimes where the two types of fields
live democratically.

\subsection{Thermo-mutual information}
\label{sec:TMI}

We can now define the thermo-mutual information (TMI).
Consider two subsets $A,B \subset \Sigma_t$ and the associated 
sub-algebras of observables $\cA_1(A) \in \cA_1(\cM)$ and 
$\cA_2(B) \in \cA_2(\cM)$. 
Using these ingredients we may construct the reduced density matrices
\eqn{
  \rho_{A1} &:=& \rho(\Omega; \cA_1(A)) \\
  \rho_{B2} &:=& \rho(\Omega; \cA_2(B)) \\
  \rho_{A1 \cup B2} &:=& \rho(\Omega; \cA_1(A) \cup \cA_2(B)) .
}
Since $\cH_i$ may be factorized in position space we may
decompose $\cH_1 = \cH_{A1} \otimes \cH_{\overline{A}1}$, etc;
then these density matrices are simply
\eqn{
  \rho_{A1} &=& \tr_{\cH_{\overline{A}1}} \tr_{\cH_2} \rho ,\\
  \rho_{B2} &=& \tr_{\cH_1}\tr_{\cH_{\overline{B}2}} \rho ,\\
  \rho_{A1 \cup B2} &=& \tr_{\cH_{\overline{A}1}} 
  \tr_{\cH_{\overline{B}2}} \rho .
}
The first two density matrices correspond to reduced density 
matrices of the type considered in \S\ref{sec:review}; the third 
is a hybrid. We define the thermo-mutual information to be:
\eq{ \label{eq:TMI}
  \TMI(A_1 : B_2) :=
  S(\rho_{A1}) + S(\rho_{B2}) - S(\rho_{A1 \cup B2}) .
}
TMI has the same structure as MI but we reserve the notation 
MI to refer to the case where the two subalgebras 
$\cA_i(A)$, $\cA_j(B)$ belong to the same sector $i=j$.
We could equivalently define TMI with the structure of the 
function $F(A : B)$ (\ref{eq:F}) as the last term
$S(\rho_{A1 \cap B2})$ vanishes due to the fact that
$\cA_1(A)\cap \cA_2(B) = 0$.

Thermo-Mutual information is indeed a kind of mutual information.
In particular:
\begin{enumerate}[i)]
  \item \label{pt:TMI_nonneg}{\bf TMI is non-negative:}  
    Since $\cA_1(A)$ and $\cA_2(B)$ are independent their associated 
    density matrices satisfy strong subadditivity (\ref{eq:SSA}) and thus
    \eq{
      \TMI(A_1 : B_2) \ge 0 .
    }
  \item \label{pt:TMI_correlators} {\bf TMI bounds local correlators 
      between $A$ and $B$:}
    This is once again a result of the Pinsker inequality. 
    For $\cO_{A1} \in \cA_1(A)$ and $\cO_{B2} \in \cA_2(B)$ the
    Pinsker inequality yields the bound
    \eq{ \label{eq:TMIPinsker}
      I(A_1 : B_2) \ge \half \left(
        \frac{\C{\cO_{A1} \cO_{B2}}_\Omega 
          - \C{\cO_{A1}}_\Omega \C{\cO_{B2}}_\Omega}
        {||\cO_{A1}|| ||\cO_{B2}||}\right)^2 .
    }
\end{enumerate}
Other salient features of TMI are:
\begin{enumerate}[i)]
  \setcounter{enumi}{2}
  \item \label{pt:TMI_lowerbound} {\bf TMI provides a lower bound for MI}
    When $A$ and $B$ are disjoint the following inequality holds:
    \eq{
      \MI(A_1 : B_1) - \TMI(A_1 : B_2) 
      = S(A_1 \cup B_2) - S(A_1 \cup B_1) \ge 0 .
    }
    This is a simple consequence of the fact that the tensor product
    $\cA_1(A) \otimes \cA_2(B)$ is always equally or more disordered
    than $\cA_1(A) \otimes \cA_1(B)$. While we do not have an explicit proof of this, it holds in all cases we study, and likely follows simply from subadditivity and the explicit form of the thermo-field double state.
  \item \label{pt:TMI_UV} {\bf TMI is free of UV singularities:}
    Just as with mutual information and the F function, in the limit $
    a\to 0$ the
    entanglement entropies in (\ref{eq:TMI}) have UV divergences.
    So long as the UV divergences are indeed universal and 
    geometric as described in \S\ref{sec:review}, these divergences
    cancel within 
    $\TMI(A_1 : B_2)$ for all subsets $A$ and $B$, even when $A \cap B \neq 0$.
    To see this note the following two facts. First, the
    ``number of effective degrees of freedom'' in the
    type-1 and type-2 sectors are equal. Second, because $\cA(A_1)$
    and $\cA(B_2)$ are independent the geometric region relevant to 
    the divergent terms in $S(\rho_{A1 \cup B2})$ is 
    ${\rm vol}(\d A + \d B)$ rather than 
    ${\rm vol}(\d (A \cup B))$ as in the case of MI.

  \item \label{pt:TMI_upperbound} {\bf TMI is bounded above for
      $\beta > 0$:}
    This follows from (\ref{pt:TMI_lowerbound}) and (\ref{pt:TMI_UV})
    provided that the theory does not suffer divergences
    associated to large separations (infrared divergences).
    The $\beta \to 0$ limit is discussed below.

  \item \label{pt:TMI_positive} {\bf TMI is non-zero for some $A$, $B$:}
    This is a trivial corollary of (\ref{pt:TMI_correlators}),
    and is necessary for $\Omega$ to be entangled.
    Assuming $0 < \beta < \infty$ 
    then $\Omega$ does not factorize into a product state on $\cH_{\rm TFD}$
    Assuming $\cA(\cM)$ is non-empty, then there exists an observable
    $\cO$ such that $\C{\cO_1 \cO_2}_\Omega > 0$.

\end{enumerate}

\noindent We also note the asymptotic behavior of TMI as $\beta \to \infty$ and $\beta \to 0$:

\begin{enumerate}[i)]
  \setcounter{enumi}{6}

  \item \label{pt:TMI_betaToInfty} 
    {\bf TMI vanishes as $\beta \to \infty$ (temperature $T \to 0$):}
    In this limit
    the canonical ensemble becomes a pure state of lowest energy;
    equivalently, in the TFD formalism
    $\Omega \to \ket{E_0}_1\otimes \ket{E_0}_2$.
    Pure states have maximal correlation, so in this limit the 
    mutual information tends to its maximum value. In contrast,
    the TMI tends to zero in this limit because the type-1 and type-2
    sectors are no longer entangled.

  \item \label{pt:TMI_betaTo0} {\bf The $\beta \to 0$ ($T\to\infty$) limit:}
    In this limit the canonical ensemble becomes random and there are 
    no (connected, normalized) correlations. For disjoint regions both the 
    MI and TMI tend to zero (with the MI bounding the TMI from above).
    When the regions overlap we generically find that the TMI diverges 
    as $\beta \to 0$. This does not contradict the points above.

\end{enumerate}

We may summarize these points by stating that TMI provides a robust
measure of the correlation between type-1 and type-2 fields.
Two important roles of the $\TMI(A_1 : B_2)$ are that it bounds 
from below the $\MI(A : B)$, and it bounds from above the 
correlation between any pair of 
observables in $\cA_1(A)$ and $\cA_2(B)$.

\subsection{Example in quantum mechanics: the two-spin system}
\label{sec:qm}

As a first introduction to TMI let us consider an example in the
simple setting of quantum mechanics. Consider a system of two
``sites'' each with a two-spin degree of freedom governed by the 
Hamiltonian
\eq{
  H = \vec{S}_A \cdot \vec{S}_B .
}
The sites are labelled $A$ and $B$ and the spin vector
$\vec{S_A} = S_A^x + S_A^y + S_A^z$ and likewise for $\vec{S}_B$.
More details than one could ever want to know about this system 
can be found in any good quantum mechanics textbook 
(e.g., Ch.~3 of \cite{Sakurai:1993aa}).
Each site has two states which we label by 
the eigenvalues $\pm$ of $S_A^z, S_B^z$, and so
the two-site system has four tensor-product states. The energy
eigenkets of the system are\footnote{The energy eigenkets correspond to the simultaneous
  eigenkets of $(\vec{S}_A + \vec{S}_B)^2$ and $(S_A^z + S_B^z)$.
}
\eqn{
  \begin{array}{ll}
  \ket{\a_1} = \frac{1}{\sqrt{2}}
  \Big( \ket{+}_A\otimes\ket{-}_B - \ket{-}_A\otimes\ket{+}_B\Big) ,
  & \quad E_1 = -3 , \\
  \ket{\a_2} = \frac{1}{\sqrt{2}}
  \Big( \ket{+}_A\otimes\ket{-}_B + \ket{-}_A\otimes\ket{+}_B\Big) ,
  & \quad E_2 = 1 , \\
  \ket{\a_3} = \ket{+}_A\otimes\ket{+}_B ,
  & \quad E_3 = 1 , \\
  \ket{\a_4} = \ket{-}_A\otimes\ket{-}_B ,
  & \quad E_4 = 1 .
  \end{array}
}

We first consider the system as a canonical ensemble.
The density matrix describing the thermal state $\Omega$ with
inverse temperature $\beta$ is
\eq{
  \rho = \frac{e^{-\beta H}}{Z} 
  = \sum_{n=1}^4 \frac{e^{-\beta E_n}}{Z} \ket{\a_n}\bra{\a_n} ,
  \quad 
  Z = \sum_{n=1}^4 e^{-\beta E_n} = e^{3\beta} + 3e^{-\beta} .
}
The entanglement entropy of the entire system is just the 
thermodynamic entropy which is easily computed:
\eq{\label{eq:qm_S}
  S(\rho) = \frac{1}{3+e^{4\beta}}\left\{
    3 \log(3+e^{4\beta}) 
    - e^{4\beta} \log\left(1-\frac{3}{3+e^{4\beta}}\right)
    \right\} .
}
If we trace of one of the sites then we are left with a random
ensemble; the reduced density matrix is (in the spin basis)
is $\rho_A = {\rm diag}\{1/2,1/2\}$. The entanglement entropy
of a single site is therefore
\eq{ \label{eq:qm_SA}
  S(\rho_A) = S(\rho_B) = \log 2 .
}
From (\ref{eq:qm_S}) and (\ref{eq:qm_SA}) we obtain the mutual
information $\MI(A : B) = 2\log2 - S(\rho)$.

Now let us apply the TFD formalism to the system. The state $\Omega$
becomes an entangled pure state
\eq{
  \ket{\Omega} = \frac{1}{\sqrt{Z}} \sum_{n=1}^4 e^{-\beta E_n}
  \ket{\a_n}_1 \otimes \ket{\a_n}_2, 
}
To compute the TMI we need the density
matrix for $A_1 \cup B_2$; it is a straightforward if aggravating
exercise to trace over the $B_1$ and $A_2$ spin sites in order
to obtain this density matrix. This matrix is not diagonal in
the $A_1 \otimes B_2$ spin basis, but is easily diagonalized.
The entanglement entropy may then be computed:
\eqn{ \label{eq:qm_SA1B2}
  S(\rho_{A1 \cup B2}) &=& -\frac{1}{4(3+e^{4\beta})}
  \Bigg\{
  (-3+e^{2\beta})^2 
    \log\left[\frac{(-3+e^{2\beta})^2 }{4(3+e^{4\beta})}\right]
    \nn \\ & & \phantom{-\frac{1}{4(3+e^{4\beta})} \Bigg\{}
    + 3 (1+e^{2\beta})^2 
    \log\left[
      \frac{\cosh^2(\beta)}{4\cosh(2\beta)-2\sinh(2\beta)}
      \right] \bigg\} ,
}
and from this we obtain $\TMI(A_1 : B_2) = 2 \log 2 - S(\rho_{A1 \cup B2})$.

\begin{figure}[h!]
  \centering
  \includegraphics[]{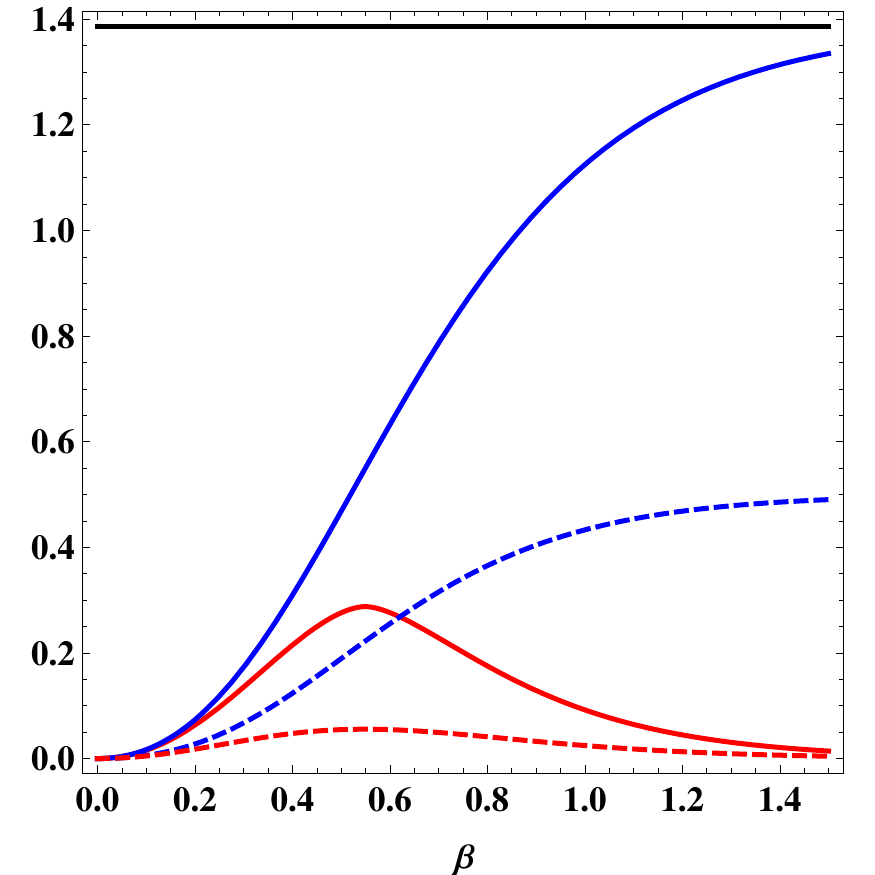}
  \caption{The mutual information and thermo-mutual information
    in the two-spin system. The solid blue line is the MI and the 
    solid red line is the TMI. The solid black line at the top of the
    graph denotes $2\log 2$ which is the maximum 
    value of the mutual information. The dashed blue and red lines
    denote the correlation functions
    $\half \left(\C{S_{A1}^z S_{B1}^z}_\Omega\right)^2$
    and
    $\half \left(\C{S_{A1}^z S_{B2}^z}_\Omega \right)^2$
    respectively.
    \label{fig:qm}}
\end{figure}

We plot $\MI(A_1 : B_1)$ and $\TMI(A_1 : B_2)$
as a function of $\beta$ in Fig.~\ref{fig:qm}.
All features of this plot are as anticipated:
\begin{enumerate}[i)]
\item
  The TMI bounds the MI from below; this may be analytically verified
  by examining the quantity
  \eq{
    \MI(A_1 : B_1) - \TMI(A_1 : B_2)
    = - S(\rho) + S(\rho_{A1 \cup B2}) \ge 0 ,
  }
  with $S(\rho)$ and $S(\rho_{A1 \cup B2})$ as in (\ref{eq:qm_SA})
  and (\ref{eq:qm_SA1B2}).
  \item
    Also plotted in Fig.~\ref{fig:qm} 
    are the right-hand sides of the Pinsker inequalities
    (\ref{eq:IPinsker}) and (\ref{eq:TMIPinsker}) for the observables
    $S_{A1}^z S_{B1}^z$ and $S_{A1}^z S_{B2}^z$. For the case at hand 
    these are simply
    \eq{
      \half \left(\C{S_{A1}^z S_{B1}^z}_\Omega\right)^2
      = \frac{(e^{4\beta}-1)^2}{2(e^{4\beta}+3)^2} ,
      \quad
      \half \left(\C{S_{A1}^z S_{B2}^z}_\Omega \right)^2
      = \frac{2(e^{2\beta}-1)^2}{(e^{4\beta}+3)^2} .
    }
    These are examples of observables bounded above by the 
    MI and TMI respectively.

  \item
    In the $\beta \to 0$ limit both the MI and TMI tend to $0$
    (recall point (\ref{pt:TMI_betaTo0}) of \S\ref{sec:TMI}).
  \item
    In the $\beta \to \infty$ limit the mutual information tends 
    to its maximal value of $2 \log 2$ while the TMI tends to zero
    (recall point (\ref{pt:TMI_betaToInfty}) of \S\ref{sec:TMI}).
\end{enumerate}

\subsection{Computing TMI in field theory}
\label{sec:QFT}

The computation of entanglement entropy in field theories is considerably 
more involved than in the simple quantum mechanical example above.
The standard prescription known as the ``replica trick'' involves
the following steps (see, e.g., \cite{Calabrese:2004eu}). One first
constructs a path integral representation of the reduced density matrix 
$\rho_A := \rho(\Psi;\cA(A))$. Using this representation one may compute the 
``R\'enyi moments'' $\tr_{\cH_A}(\rho_A^n)$.
This procedure amounts to gluing together path integral 
representations such that $\tr_{\cH_A}(\rho_A^n)$ 
maybe be viewed as a single path integral over a Riemann surface
rather than the original space $\cM$.
Assuming the expression for $\tr_{\cH_A}(\rho_A^n)$ is a complex 
analytic function of $n$ in a connected region $R$ of the complex $n$ plane
which includes the non-negative integers, one may 
obtain the entanglement entropy $S(\rho_A)$ via
\eq{
  S(\rho_A) = - \tr_{\cH_A} \rho_A \log \rho_A
  = - \lim_{n\to 1} \frac{\d}{\d n} \tr_{\cH_A} (\rho_A^n) .
}
The path integrals involved in this prescription are on most solid 
footing in Euclidean signature.

\begin{figure}[t!]
  \centering
  \includegraphics[width=0.5\textwidth]{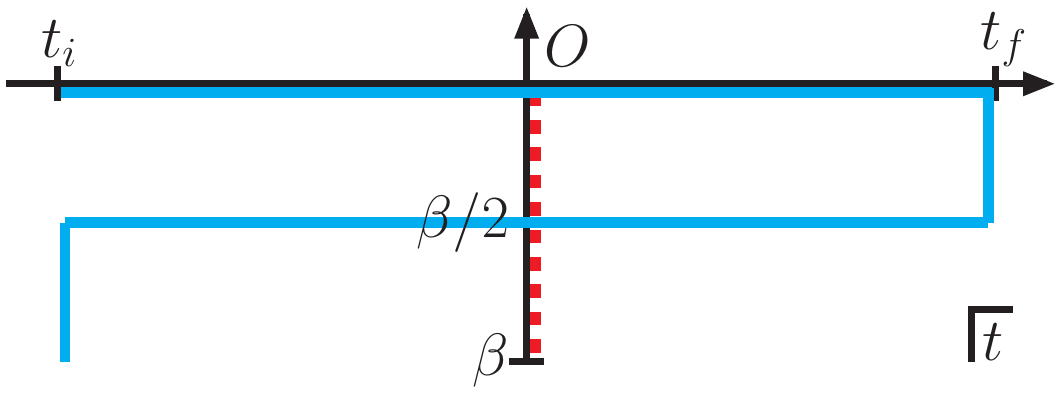}
  \caption{Schwinger-Keldysh time integration contours in the 
    complex $t$ plane. The Euclidean contour is the dashed red line; 
    the TFD contour is the solid blue line. These contours are further
    discussed in Appendix~\ref{app:PI}.
    \label{fig:SKcontour}
  }
\end{figure}

In order to utilize the replica trick to construct both the MI and TMI
we need a path integral representation of the reduced density 
matrices $\rho_{A1}$, $\rho_{B1}$, $\rho_{B2}$, $\rho_{A1 \cup B1}$, and 
$\rho_{A1 \cup B2}$. Life is simpler if we work in Euclidean
signature $\cM \to \cM_{\rm E} = S^1\times \Lambda$ where the
Euclidean time circle has radius $\beta$. The only subtlety that arises
is understanding how type-1 and type-2 fields are distinguished in
Euclidean signature. Recall that the Euclidean and TFD formulations
of thermal field theory are related by a deformation of the 
Schwinger-Keldysh time contour -- see Fig.~\ref{fig:SKcontour}.
From this perspective there are two types of operators in Lorentz
signature because the Lorentzian time contour has two legs. However, there 
is just one leg to the Euclidean time contour and therefore only one set 
of Euclidean operators. 
So care must be taken to define the Euclidean analogue of a density
matrix in which one type of Lorentzian operators has been traced over.
At the end of the day, the Euclidean path integrals defining these density
matrices are an intuitive generalization of existing results
\cite{Calabrese:2004eu,Calabrese:2009qy}, so we simply quote the
results below and relegate a careful derivation to Appendix~\ref{app:PI}.

\begin{figure}[h!]
  \centering
  \includegraphics[width=\textwidth]{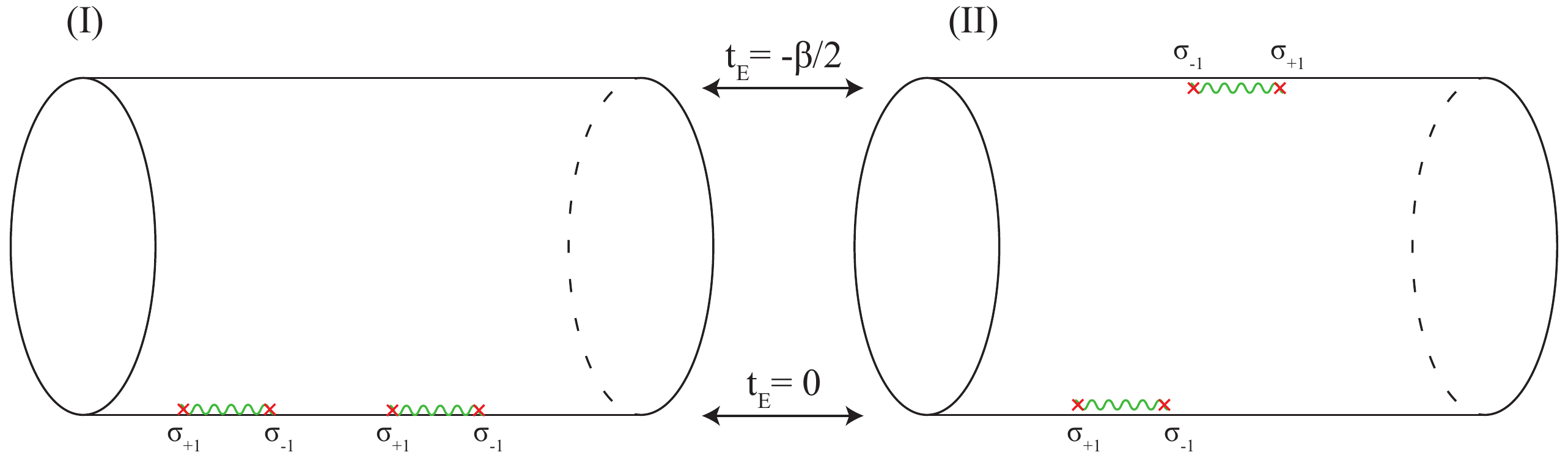}
  \caption{ Depiction of the Euclidean path integral (\ref{eq:rhoABE}).
    (I) shows the placement of delta function insertions used
    to construct the Euclidean analogue of $\rho_{A1 \cup B1}$;
    (II) shows the placement for the Euclidean analogue of 
    $\rho_{A1 \cup B2}$. The twist fields (see \S\ref{sec:CFT})
    are also depicted, note that the orientation of the cuts at $-\beta/2$ are reversed. \label{fig:cylinder}
  }
\end{figure}

We use as a coordinate on $\cM_{\rm E}$ the complex coordinate $z$ with 
$\Re z = x \in \Reals$ and $\Im z \in (-\beta,0)$. 
For $A$ a single interval the density matrix $\rho_{A1}$ has the
Euclidean path integral representation
\eqn{ \label{eq:rhoA1Euclidean}
  \rho_{A1}(\psi', \psi) &=&
  Z^{-1} \int [\cD\phi] e^{-S_{\rm E}[\phi]}
  \left[\prod_{x\in A} 
    \delta\Big(\phi(0^+)-\psi'(0^+)\Big) \delta\Big(\phi(0^-)-\psi(0^-)\Big)
    \right] .
}
Here $\psi',\psi$ denote Euclidean field profiles and we have
labelled only the dependence on $\Im z$, i.e., 
$\phi(0^+) = \phi(\Re z,\Im z = 0 +i\epsilon)$, etc.
Away from the delta function insertions the Euclidean fields $\phi(x)$
satisfy the KMS boundary conditions $\phi(z) \sim \phi(z-i\beta)$.
Eq. (\ref{eq:rhoA1Euclidean}) is precisely the path integral obtained 
from the zero-temperature
case by compactifying the Euclidean time direction \cite{Calabrese:2009qy}.
The expression for $\rho_{A1 \cup B1}$ follows suit:
\eqn{
  \rho_{A1\cup B1}(\psi', \psi) &=&
  Z^{-1} \int [\cD\phi] e^{-S_{\rm E}[\phi]} \Bigg\{
  \left[\prod_{x\in A} 
    \delta\Big(\phi(0^+)-\psi'(0^+)\Big) \delta\Big(\phi(0^-)-\psi(0^-)\Big)
    \right]
    \nn \\ & & \phantom{Z^{-1} \int [\cD\phi] \;}
  \left[\prod_{x\in B} 
    \delta\Big(\phi(0^+)-\psi'(0^+)\Big) \delta\Big(\phi(0^-)-\psi(0^-)\Big)
    \right] \Bigg\} .
}
The path integral for a type-2 density matrix differs from 
(\ref{eq:rhoA1Euclidean}) in two respects: the delta function insertions 
are placed at $\Im z = -\beta/2$, and the $i\epsilon$ prescriptions are 
reversed:
\eqn{ \label{eq:rhoA2Euclidean}
  \rho_{A2}(\psi', \psi) &=&
  Z^{-1} \int [\cD\phi] e^{-S_{\rm E}[\phi]}
  \Bigg[\prod_{x\in A} 
    \delta\Big(\phi(-\beta/2 + 0^-) -\psi'(-\beta/2+ 0^-)\Big)
    \nn \\ & & \phantom{Z^{-1} \int [\cD\phi] e^{-S_{\rm E}[\phi]}\Bigg[\; }
    \delta\Big(\phi(-\beta/2 + 0^+) -\psi(-\beta/2 + 0^+)\Big)
    \Bigg] .
}
Of course, using a coordinate transformation one may recast 
(\ref{eq:rhoA2Euclidean}) in the form of (\ref{eq:rhoA1Euclidean}), so
the R\'enyi moments of $\rho_{A1}$ and $\rho_{A2}$ are 
equivalent. These differences do matter, however, in the hybrid density
matrix:
\eqn{ \label{eq:rhoA1B2E}
  \rho_{A1\cup B2}(\psi', \psi) &=&
  Z^{-1} \int [\cD\phi] e^{-S_{\rm E}[\phi]} \Bigg\{
  \left[\prod_{x\in A} 
    \delta\Big(\phi(0^+)-\psi'(0^+)\Big) \delta\Big(\phi(0^-)-\psi(0^-)\Big)
    \right]
    \nn \\ & & \phantom{Z^{-1} \int [\cD\phi] \;}
  \Bigg[\prod_{x\in B} 
    \delta\Big(\phi(-i\beta/2 + 0^-) -\psi'(-i\beta/2+ 0^-)\Big)
    \nn \\ & & \phantom{Z^{-1} \int [\cD\phi] e^{-S_{\rm E}[\phi]}\Bigg[\; }
    \delta\Big(\phi(-i\beta/2 + 0^+) -\psi(-i\beta/2 + 0^+)\Big)
    \Bigg] \Bigg\} .
}
The placement of delta insertions is depicted in Fig.~\ref{fig:cylinder}.

Given (\ref{eq:rhoA1Euclidean})-(\ref{eq:rhoA1B2E}) it is straightforward 
to proceed with the remaining steps of the replica trick. The result is simply that we study the partition function on a surface constructed by gluing thermal cylinders together, the only difference is that not all cuts are at $\Im z=0$, and the orientation of the cuts at $\Im z = -\beta/2$ are reversed. This is a consequence of our $i\epsilon$ prescription, and is reminiscent of calculations of entanglement negativity \cite{Calabrese:2012nk,Calabrese:2012ew,Das:1997gg}, where one exchanges the ordering of twist operators. Both the TMI and negativity are extracted from a correlator $\langle \sigma_+ \sigma_- \sigma_- \sigma_+ \rangle$, however their locations on the integration contour are quite different: TMI is calculated via
\be
\langle\sigma_+(u_1)\sigma_-(v_1) \sigma_-(u_2+i\beta/2) \sigma_+(v_2+i\beta/2)\rangle_\beta\ee
whereas the ``Renyi negativities'' are extracted from
\be\langle\sigma_+(u_1)\sigma_-(v_1) \sigma_-(u_2) \sigma_+(v_2)\rangle,\ee
evaluated on $Z_n$ orbifolds for even $n$, after which we must analytically continue to $n=1$.

\subsection{Example in CFT: 2d massless Dirac fermion}
\label{sec:CFT}

For our second example we consider the $2d$ massless Dirac fermion.
This theory is sufficiently simple that, by borrowing heavily from
existing results in the literature, we may compute the MI and TMI exactly.

Following the discussion of the previous section, our main task is
to compute the R\'enyi moments $\tr_{\cH_C}(\rho_C^n)$ starting from
the expression for the Euclidean path integral for $\rho_C$.
We may use the replica trick to glue together path integrals
for $\rho_C$ in order to obtain a single path integral for 
$\tr_{\cH_C}(\rho_C^n)$ over an n-sheeted Riemann surface.
For 2D CFTs we may use a further refinement of Calabrese and Cardy
\cite{Calabrese:2009qy} who have shown that this path integral may be 
related to the correlation function of twist operators $\sigma^{\pm}_n(z)$
defined on the original space $\cM_{\rm E}$.
Let us denote by $[u,v]$ a line on $\cM_{\rm E}$ with endpoints $u,v$ and
constant imaginary part, and let $C$ be the union of $N$ disjoint 
segments $C = \cup_{i=1}^N [u_i,v_i]$. The
trace $\tr_{\cH_C}(\rho_C^n)$ is proportional to a $2N$-pt. correlation
function of twist operators:
\eq{
  \tr_{\cH_C}(\rho_C)^n \propto 
  \C{\sigma_n^\pm(u_1)\sigma_n^\pm(v_1)\dots
    \sigma_n^\pm(u_N)\sigma_n^\pm(v_N)}_\Omega .
}
The twist operators are conformal primaries with conformal weight
\eq{
  \Delta_n = \frac{c}{12}\left(n-\frac{1}{n}\right)
  = \frac{c}{12}\frac{(n-1)(n+1)}{n} ,
}
with $c$ the central charge.
The choice of which twist field $\sigma_n^\pm$ is determined by the
placement of the intervals. For the case considered by \cite{Calabrese:2009qy}
where all intervals lie on the real axis ($\Im z = 0$)
the twist correlator is
\eq{
  \tr_{\cH_C}(\rho_C)^n \propto 
  \C{\sigma_n^+(u_1)\sigma_n^-(v_1)\dots\sigma_n^+(u_N)\sigma_n^-(v_N)}_\Omega ,
 \quad \Im u_i = \Im v_i = 0 \; \forall \; i=1,\dots,N .
}
For segments lying on $\Im z = -\beta/2$ the orientation of the
twist fields is reversed. This is a consequence of the $i\epsilon$
prescriptions described in \S\ref{sec:QFT}.

In order to compute the MI and TMI we consider the three segments
\eqn{
  A_1 &=& [u_{A1},v_{A1}], \quad \Im u_{A1} = \Im v_{A1} = 0 , \\
  B_1 &=& [u_{B1},v_{B1}], \quad \Im u_{B1} = \Im v_{B1} = 0 ,\\
  B_2 &=& [u_{B2},v_{B2}], \quad \Im u_{B2} = \Im v_{B2} = -\beta/2 ,
}
The R\'enyi moments for the associated density matrices are proportional
to
\eqn{
  \tr (\rho_{A1}^n) &\propto& \C{\sigma^+_n(u_{A1})\sigma^-_n(v_{A1})}_\Omega, \\
  \tr (\rho_{B1}^n) &\propto& \C{\sigma^+_n(u_{B1})\sigma^-_n(v_{B1})}_\Omega, \\
  \tr (\rho_{B2}^n) &\propto& \C{\sigma^-_n(u_{B2})\sigma^+_n(v_{B2})}_\Omega .
}
These 2-pt. functions are completely determined by conformal symmetry
and depend on the specific model only in the value of $c$ which
determines $\Delta_n$:
\eq{
  \C{\sigma^+_n(u_1)\sigma_n^-(v_1)}_\Omega 
  = \C{\sigma^-_n(u_1)\sigma_n^+(v_1)}_\Omega
  = \left| \frac{\pi \epsilon}
    {\beta \sinh\left[\frac{\pi(u_1-v_1)}{\beta}\right]} \right|^{2\Delta_n} .
}
The R\'enyi moments of unions of the regions are proportional to
\eqn{
  \tr (\rho_{A1\cup B1}^n)
  &\propto& \C{\sigma^+_n(u_{A1})\sigma^-_n(v_{A1})
    \sigma^+_n(u_{B1})\sigma^-_n(v_{B1})}_\Omega, \\
  \tr (\rho_{A1\cup B2}^n)
  &\propto& \C{\sigma^+_n(u_{A1})\sigma^-_n(v_{A1})
    \sigma^-_n(u_{B2})\sigma^+_n(v_{B2})}_\Omega .
}
Since the 4-pt. functions are not fully determined by conformal 
symmetry the right-hand side of these proportionalities are model dependant. 
For the massless Dirac fermion\footnote{As has been clarified in \cite{Headrick:2012fk}, we will be studying the unprojected Dirac fermion theory, not the modular invariant theory where fermion number is gauged.} $c=1$ and the twist correlators are known
\cite{Casini:2005rm}:\footnote{
  For this model the correlation functions of twist fields on the complex 
  plane are given by
  \eq{
    \langle \sigma_n^+(u_1)\sigma_n^-(v_1)\ldots \sigma_n^+(u_N) 
    \sigma_n^-(v_N)\rangle_{plane}= \left|\det M \right|^{2\Delta_n},~M_{ij} 
    = \frac{\epsilon}{v_j-u_i}.
  }
  The correlation functions on the cylinder may be obtained by the 
  conformal transformation $w_{plane} = \exp[2\pi z_{cyl}/\beta]$.
}
\eqn{
  & & \C{\sigma^+_n(u_1)\sigma_n^-(v_1)\sigma_n^+(u_2)\sigma_n^-(v_2)}_\Omega
  \nn \\
  & & \phantom{\quad\quad} =
  \left| \frac
    {\pi^2\epsilon^2  \shr{\pi(u_1-u_2)}{\beta}\shr{\pi(v_1-v_2)}{\beta}}
    {\beta^2  \shr{\pi(u_1-v_1)}{\beta} \shr{\pi(u_1-v_2)}{\beta} 
      \shr{\pi(v_1-u_2)}{\beta}\shr{\pi(u_2-v_2)}{\beta} }
  \right|^{2\Delta_n} ,
}
\eqn{
  & & \C{\sigma^+_n(u_1)\sigma_n^-(v_1)\sigma_n^-(u_2)\sigma_n^+(v_2)}_\Omega
  \nn \\
  & & \phantom{\quad\quad} =
  \left| \frac
    {\pi^2\epsilon^2  \shr{\pi(u_1-v_2)}{\beta}\shr{\pi(v_1-u_2)}{\beta}}
    {\beta^2  \shr{\pi(u_1-v_1)}{\beta} \shr{\pi(u_1-u_2)}{\beta} 
      \shr{\pi(v_1-v_2)}{\beta}\shr{\pi(u_2-v_2)}{\beta} }
  \right|^{2\Delta_n} .
}

We can now quickly assemble these ingredients into the MI and TMI.
The mutual information is given by
\eqn{
  \MI(A_1 : B_1)
  &=& \lim_{n\to 1} \frac{1}{1-n} \log
  \left[
    \frac{\tr (\rho_{A1}^n) \tr (\rho_{B1}^n)}{\tr (\rho_{A1 \cup B1}^n) }
  \right] 
}
We know the R\'eyni moments up to constants of proportionality which are
independent of $\beta$. These constants provide an overall constant
to the mutual information. The correct value of this constant is fixed
by the requirement that the mutual information vanish as $\beta\to 0$.
Inserting the expressions for the twist correlators we obtain
\eqn{
  \MI(A_1 : B_1)
  &=&
  \frac{1}{3} \log\left|
    \frac{\sinh\left(\frac{\pi(u_{A1}-u_{B1})}{\beta}\right)
      \sinh\left(\frac{\pi(v_{A1}-v_{B1})}{\beta}\right)}
    {\sinh\left(\frac{\pi(u_{A1}-v_{B1})}{\beta}\right)
      \sinh\left(\frac{\pi(v_{A1}-u_{B1})}{\beta}\right)}
    \right| . \label{eq:MICFT1}
}
To clean this expression up consider without loss of generality the
configuration $u_{A1} < v_{A1} < u_{B1} < v_{B1}$; we may then adopt
the variables
\eq{ \label{eq:lengths}
  |u_{A1} - v_{A1}| = L_A, \quad   |u_{B1} - v_{B1}| = L_B, \quad 
  u_{B1} - v_{A1} = S ,
}
and rewrite (\ref{eq:MICFT1}) as
\eqn{ \label{eq:MICFT2}
  \MI(A_1 : B_1)
  &=& \frac{1}{3} \log\left[
    \frac{\sinh\left(\frac{\pi(L_A + S)}{\beta}\right)
      \sinh\left(\frac{\pi(L_B+S)}{\beta}\right)}
    {\sinh\left(\frac{\pi(L_A + L_B +S}{\beta}\right)
      \sinh\left(\frac{\pi S}{\beta}\right)}
    \right] .
}
For the TMI we similarly compute
\eqn{
  \TMI(A_1 : B_2) &=&
   \frac{1}{3} \log\left|
    \frac{\sinh\left(\frac{\pi(u_{A1}-v_{B2})}{\beta}\right)
      \sinh\left(\frac{\pi(v_{A1}-u_{B2})}{\beta}\right)}
    {\sinh\left(\frac{\pi(u_{A1}-u_{B2})}{\beta}\right)
      \sinh\left(\frac{\pi(v_{A1}-v_{B2})}{\beta}\right)}
    \right| \\
    &=&
  \frac{1}{3} \log\left[
    \frac{\cosh\left(\frac{\pi(L_A+L_B+S)}{\beta}\right)
      \cosh\left(\frac{\pi S}{\beta}\right)}
    {\cosh\left(\frac{\pi(L_A+S)}{\beta}\right)
      \cosh\left(\frac{\pi(L_B+A)}{\beta}\right)}
    \right] 
}
In the second equality we have once again inserted the values 
(\ref{eq:lengths}), along with the fact that 
$\Im u_{B2} = \Im v_{B2} = -\beta/2$.
Once again the overall constant is established by demanding that
the TMI vanish as $\beta\to 0$.

\begin{figure}[h!]
  \centering
  \includegraphics[width=0.45\textwidth]{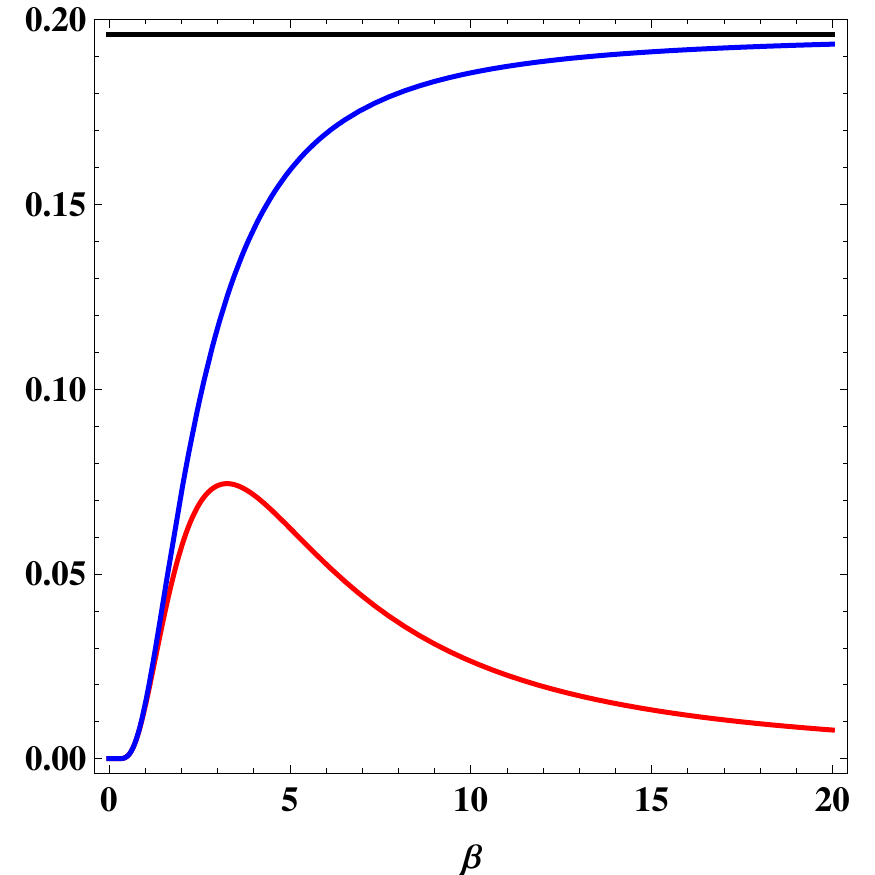}%
  \includegraphics[width=0.45\textwidth]{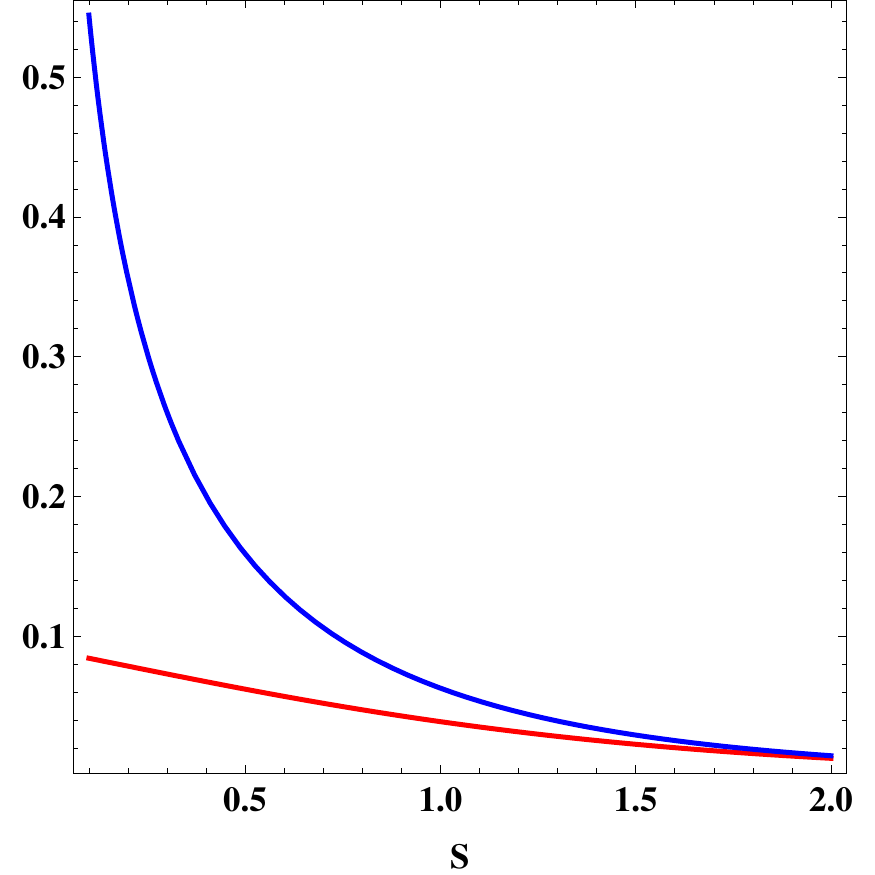}
  \caption{Examples of the MI and TMI of the 2D massless Dirac fermion. 
    Left: the MI (solid blue) and TMI (solid red) as a function of $\beta$. 
    In this plot $L_A = L_B = 1$ and $S=1/2$. The solid black line 
    denotes the maximum value of the MI.
    Right: the MI and TMI as function of separation $S$. Here 
    $L_A = L_B = 1$ and $\beta = 5$. \label{fig:CFTExample}
  }
\end{figure}

We provide representative plots of the MI and TMI in 
Fig.~\ref{fig:CFTExample}. As expected, the MI bounds the TMI from 
above. Both MI and TMI vanish as $\beta \to 0$; as $\beta \to \infty$ 
the MI approaches is maximum value while the TMI once again vanishes.

\section{Holographic thermo-mutual information}
\label{sec:HTMI}

In this section we change gears and utilize the holographic techniques of 
the AdS/CFT correspondence \cite{Aharony:1999ti} to study TMI from a 
different perspective.
In particular, we use the holographic formula of Ryu and Takayanagi
\cite{Ryu:2006bv} to examine TMI for the $2d$ CFT
dual to the non-rotating BTZ black hole in $3$ bulk dimensions.

The original RT formula \cite{Ryu:2006bv} provides a
holographic prescription for calculating entanglement entropy in a  
strongly-coupled $d$-dimensional CFT on a static spacetime $\cM$
which admits a dual description, in the usual sense of AdS/CFT, as a 
static $d+1$-dimensional asymptotically AdS spacetime
$\cB = \cM \times \Reals$. 
Because both bulk and boundary admit a timelike Killing vector
$\partial_t$, we may use the canonical foliation into equal $t$ surfaces
$\Sigma_t \subset \cB$ and $\partial\Sigma_t \subset \cM$.
Consider a region $A \subset \partial \Sigma_t$ on the boundary. 
The RT proposal is that the entanglement entropy $S_A$ is given 
by\footnote{We are working in Einstein frame with the action 
normalized as $S=\frac{1}{16\pi G_N}\int\sqrt{-g}R+\ldots$ and ignoring 
higher-derivative corrections.}
\be
  S_A=\frac{1}{4G_N} \min_{M_A}\left[ \mathrm{Vol}(M_A)\right],
  \label{entanglement_formula}
\ee
where $M_A \subset \Sigma_t$ is a surface in the bulk geometry homologous 
to $A$.
While this is only a conjectured formula, it has passed many nontrivial checks (see e.g. \cite{Nishioka:2009un,Headrick:2010zt,Hayden:2011ag} for a review). When the system in question is no longer static (\ref{entanglement_formula}) must be generalized to the covariant holographic entanglement formula of \cite{Hubeny:2007xt}. The clearest presentation of this generalization replaces minimal surfaces with saddlepoints of the area action for spacelike co-dimension two bulk surfaces homologous to $A$. In $3$ bulk geometries these surfaces are 
simply spacelike geodesics, and the entanglement entropy is then again 
given by (\ref{entanglement_formula}).

\begin{figure}
  \centering
  \includegraphics[width=0.4\textwidth]{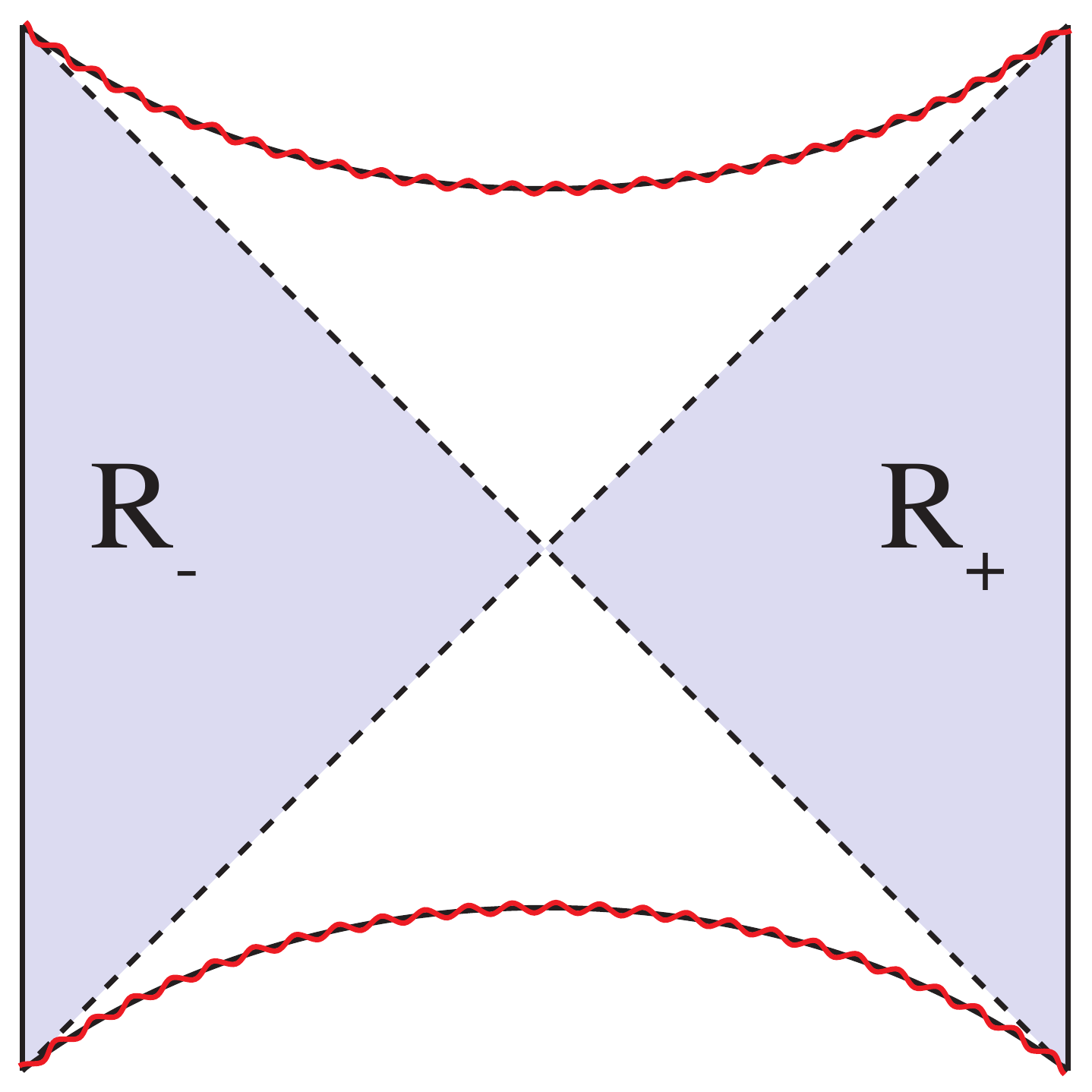}
  \caption{The Penrose-Carter diagram
    for a maximally extended AdS-Schwarzschild black hole. The two dark shaded regions labeled $R_\pm$ are the regions covered by two AdS-Schwarzschild coordinate patches where the killing vector $\partial_t$ is timelike, bounded by the $45^\circ$ dashed lines indicating the horizon. The red jagged line is the curvature singularity, which bends inwards for AdS-Schwarzschild black holes in $D>3$.
    \label{fig:penrose}
    }
\end{figure}

In AdS/CFT thermal systems at high enough temperature
and without chemical potential are described by 
non-rotating eternal black holes. These manifolds have two exterior regions
outside the black hole horizon and correspondingly two conformal boundaries, 
see figure \ref{fig:penrose}.
While these manifolds do not admit a global timelike Killing vector field,
they do admit Killing vector fields which are timelike in the exterior
regions.
The AdS/CFT interpretation in this setting was established by Maldacena 
\cite{Maldacena:2001kr} and in many way follows the classic work of Israel 
\cite{Israel:1976ur}. The two boundaries correspond to the two copies
of the CFT in the TFD description of the thermal state. 
Due to the presence of timelike Killing vector fields on the boundaries
the CFT is governed by a time-independent Hamiltonian.
The entanglement 
entropy of the CFT state is realized geometrically as the entropy of the 
black hole.
Basic properties of the TFD system described \S\ref{sec:thermal},
such as the fact that all type-1 and type-2 operators commute, are
clearly realized through this bulk description.

Several authors have used the RT formula to study mutual information in 
thermal systems holographically 
(see e.g., \cite{Hubeny:2007xt,Headrick:2010zt}). In order to distinguish
the holographic prediction from exact results in the field theory
we refer to this as the holographic mutual information (HMI).
Let $\partial_t$ be the
timelike Killing vector field in a exterior region, and consider
two regions $A_1,B_1 \subset \partial\Sigma_t$ on the
conformal boundary of this exterior region. 
The HMI between $A_1$ and $B_1$ is simply
\eq{
  \HMI(A_1 : B_1) := S_{A1} + S_{B1} - S_{A1\cup B1} ,
}
with the holographic entanglement entropies computed from the 
RT formula (\ref{entanglement_formula}).
The entropies $S_{A1}$ and $S_{B1}$ are given by $\vol(M_{A1})$
and $\vol(M_{B1})$, respectively, where $M_{A1,B1} \subset \Sigma_t$ are 
the minimal surfaces in the bulk homologous to $A_1$ and $B_1$.
For the entropy $S_{A1\cup B1}$ there are two candidate minimal surfaces:
i) the disconnected surface $M_{A1} \cup M_{B1}$, and ii) the
connected minimal surface $M_C$ homologous to $A_1 \cup B_2$.
The entropy $S_{A1\cup B1}$ is determined by the lesser volume of
these two choices. Thus we may compactly write the HMI as 
\eq{
  \HMI(A_1 : B_1) = \frac{1}{4 G_N}
  \max \left[ \vol(M_{A1}) + \vol(M_{B1}) - \vol(M_C), 0 \right] ,
}
where the zero result is obtained when the $S_{A1\cup B1}$ is given
by the disconnected minimal surface.
It is important to note that none of these surfaces cross the 
black hole horizon; the geometry behind 
the horizon is not needed for (or probed by) this computation. 
This is the analogue of the fact that mutual information may can be 
constructed from only one copy of the CFT -- the second copy is traced 
out to reproduce the thermal density matrix.

The holographic TMI may be computed in a similar manner to the HMI.
Since thermo-mutual information involves operators of both type-1 and type-2,
the holographic thermo-mutual information (HTMI) involves 
regions on each conformal boundary of the extended black hole spacetime. 
Extended black hole spacetimes do not admit a global timelike Killing vector 
field, so in general the entanglement entropies must be computed using the 
covariant form of the RT proposal. 
As already mentioned, however, in $3$ bulk dimensions this does not 
affect the computation.
Let $A_1, B_2 \in \partial\Sigma_t$ be two regions on \emph{different}
conformal boundaries.
The HTMI between $A_1$ and $B_2$ given by
\eq{
  \HTMI(A_1 : B_2) = \frac{1}{4 G_N}
  \max \left[ \vol(M_{A1}) + \vol(M_{B1}) - \vol(M_C), 0 \right] ,
}
where now $M_C$ is the the minimal connected surface connecting $A_1$
to $B_2$ which in addition satisfies a regularity condition we describe
momentarily.
Because there is an Einstein-Rosen bridge 
connecting the two asymptotic regions, there do indeed exist co-dimension 
two surfaces which connect $A_1$ to $B_2$.
In general, finding these minimal surfaces for maximally extended black holes 
analytically can be difficult; however, in $3$ dimensions the computation 
is much simpler because BTZ black holes are quotients of $AdS_3$.

The regularity constraint we impose on $M_C$ in the computation
of HTMI is the Lorentzian analog of the homology constraint imposed in the RT
formula. The homotopy constraint says that, for the geodesics corresponding to $A_1\cup B_2$, there exists a \emph{spacelike} co-dimension one surface $r_{A1\cup B2}$,  in the bulk such that $\partial r = A_1 \cup B_2 \cup M_{A1\cup B2}$. This requirement that the surface is space like means that in $AdS_3$ the ribbon between the geodesics can not twist around, as it would no longer be spacelike. This condition is crucial, as there are configurations where the geodesics corresponding to a twisted ribbon are shorter than the untwisted case (for instance, regions with angular width $\sim\pi$ on opposite sides of the circle) but are rejected due to failing the homology constraint. 

 We are also not interested in surfaces which wrap the spatial $S^1$ multiple times - these are guaranteed to be longer than the untwisted ``straight across'' geodesic, and so can be ignored at leading order.

Before proceding to the compuational details, let us briefly
remind the reader of the Hawking-Page \cite{Hawking:1983aa} phase transition 
that occurs for AdS black holes at low temperature ($\beta \approx 2\pi$ 
in conventions defined below). For simplicity we focus on the 
$2+1$-dimensional case.
At high temperatures $\beta \lesssim 2\pi$ the dominant saddle point
of the gravity path integral (with asymptotically AdS boundary conditions)
is given by the BTZ black hole. In contrast, at low temperature 
$\beta \gtrsim 2 \pi$ the dominant saddlepoint is global AdS. This 
can be infered, for instance, by comparing the thermodynamic free energy 
$F = E - T S$ of the two spacetimes:
\eq{
  \Delta F = F_{BTZ} - F_{AdS} 
  = -\frac{\pi^2\ell^2}{2 G_N\beta^2} + \frac{1}{8 G_N} .
}
From this expression one determines that a first-order phase transition 
occurs at $\beta =2\pi \ell$.
Obviously, the relevant spacetime for use in the RT formula is the
dominant saddlepoint.  It is clear that in the $AdS$ phase the saddlepoint is two copies of global AdS, with no Einstein-Rosen bridge connecting them, and therefore the TMI will always vanish. Below we focus attention on the high-temperature 
regime where BTZ provides the correct gravity solution.

\subsection{BTZ geometry}
\label{sec:BTZ}

This section serves as a brief review of the BTZ geometry.
We emphasize the description of BTZ in terms of an embedding space
as this formulation makes our computations below quite simple.
Similar treatments can be found in \cite{Kraus:2002iv,Fidkowski:2004aa}.

\begin{figure}
  \centering
  \includegraphics[width=0.4\textwidth]{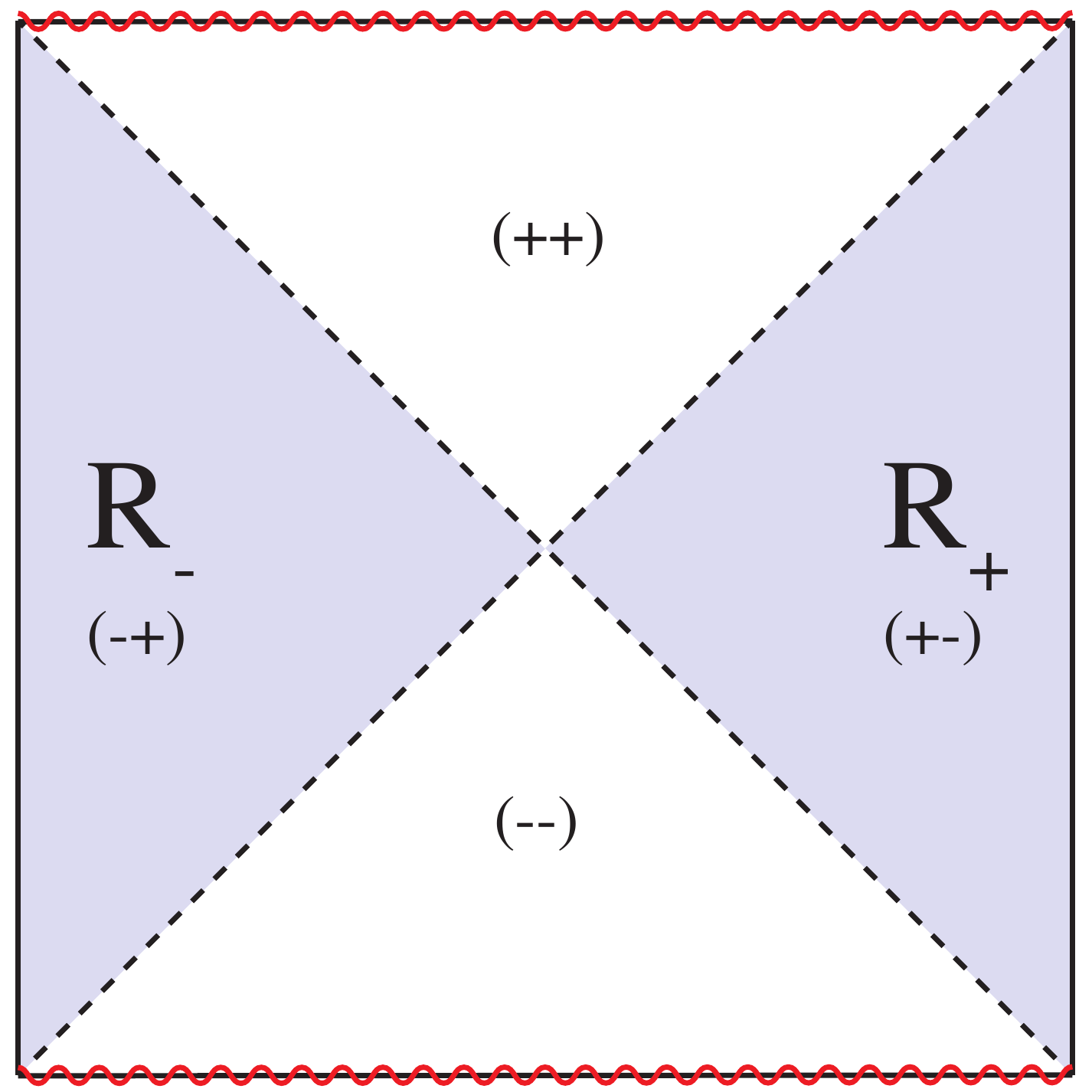}
  \caption{The Penrose-Carter diagram
     of BTZ.  The darker shaded regions  are $R_+$
    and  $R_-$, covered by the AdS-Schwarzschild patches (\ref{eq:pmw}). Note that unlike higher-dimensional black holes, the singularity is not bent in.
    \label{fig:btz}The $(\cdot,\cdot)$ indicate the signs of $w_+$ and $w_-$ in each region.
    }
\end{figure}

Recall that $AdS_3$ may be constructed as a 3-dimensional surface 
in a 4-dimensional embedding space $\Reals^{2,2}$. Adopting the metric 
$g_{AB} = \text{diag}\{-,+,+,-\}$, for this space, $AdS_3$ is the 
universal cover of the hypersurface defined by
\eq{ \label{eq:surface}
  - \ell^2 = g_{AB} X^A X^B
  = - (X^0)^2 + (X^1)^2 + (X^2)^2 - (X^3)^2 ,
}
with $\ell$ the AdS radius. 
BTZ is a quotient of $AdS_3$ and also admits a simple
description as an embedded surface. 
If we identify $X^2 \pm X^3 \cong e^{\pm 4 \pi^2 / \beta} (X^2 \pm X^3)$
in the embedding space $\Reals^{2,2}$ then the BTZ geometry is given 
by the hypersurface (\ref{eq:surface}). We obtain the induced metric on the BTZ geometry by introducing
the dimensionless real coordinates $(w_+,w_-,\phi)$ via
\eqn{
  X^0 &=& \ell \left(\frac{-w_+ - w_-}{1+w_+w_-}\right) , \nn \\
  X^1 &=& \ell \left(\frac{-w_+ + w_-}{1+w_+w_-}\right) , \nn \\
  X^2 &=& \ell \left(\frac{1 - w_+w_-}{1+w_+w_-}\right) \sinh \phi , \nn \\
  X^3 &=& \ell \left(\frac{1 - w_+w_-}{1+w_+w_-}\right) \cosh \phi .
}
These coordinates have the respective ranges:
\eq{
  w_+\in\Reals,\quad w_-\in\Reals,\quad -1 < w_+ w_- < 1,\quad
  \phi\cong\phi+\frac{4\pi^2}{\beta}.
}
The BTZ line element is
\eq{ \label{eq:wmetric}
  ds^2 = \frac{\ell^2}{(1+w_+ w_-)^2}
  \left[ -4 dw_+ dw_- + (-1+w_+w_-)^2 d\phi^2 \right] .
}
Clearly $w_+$ and $w_-$ are null coordinates and $\phi$ is an angular
coordinate. There is a conformal boundary at $w_+w_- = -1$, a horizon
at $w_+ w_- = 0$, and a conical singularity at $w_+w_- = +1$. See 
Fig.~\ref{fig:btz}.

For our purposes it is more useful to employ AdS-Schwarzschild
coordinates. Consider the regions 
$R_+ := \{0\le w_+, -1\le w_+ w_- \le 0\}$ and
$R_- :=  \{w_+\le 0, -1\le w_+ w_- \le 0\}$ respectively; 
each region is outside the horizon and may be covered by a
copy of the AdS-Schwarzschild chart
\eqn{
  z &=& z_H\frac{1+w_+w_-}{1-w_+w_-}, \quad z \in (0,z_H) ,
  \quad z_H = \frac{\beta}{2\pi} , \nn \\
  t &=& \frac{z_H}{2} \ln\left(-\frac{w_+}{w_-}\right) , \quad t\in \Reals , \nn \\
  \theta & = & z_H \phi,\quad \theta \cong \theta+2\pi, \nn \\
  w_+ &=& \pm \left(\frac{z_H-z}{z_H+z}\right)^{1/2} e^{t/z_H}, \quad
  w_- = \mp \left(\frac{z_H-z}{z_H+z}\right)^{1/2} e^{-t/z_H}, 
  \label{eq:pmw}
}
\eq{ \label{eq:SchwmetricDimensionful}
  ds^2 = \frac{\ell^2}{z^2} \left[ 
    - \left(1-\frac{z^2}{z_H^2}\right) d t^2
    + \left(1-\frac{z^2}{z_H^2}\right)^{-1} dz^2
    + d\theta^2 \right] .
}
The choice of sign in the last line of (\ref{eq:pmw}) is $\pm$ for
$R_\pm$. We can analytically continue from $R_+$ to $R_-$ via
$t\rightarrow t-i\beta/2 $.
In these coordinates the conformal boundary is at $z = 0$ and the horizon
is at $z = z_H$. Using this normalization of time, we see that the horizon 
has a temperature $\beta^{-1}$. 
The full BTZ geometry also has ``upper'' and ``lower'' wedges 
(recall Fig.~\ref{fig:btz}); in  these regions $\d_t$ is spacelike.

The extended manifold does not enjoy a global timelike isometry, but
it does admit a boost isometry that is timelike in the regions $R_{\pm}$;
the boost vector is
\be
\frac{(1+w_+ w_-)^2}{4 w_- }\d_+- \frac{(1+w_+ w_-)^2}{4 w_+ }\d_-=\frac{\ell^2}{z_H}\d_t ,
\ee
which has opposite orientation in the two regions $R_\pm$. 
Kruskal time $T=\frac{w_++w_-}{2}$ provides a global notion of time but $\d_T$ 
is not an isometry.

The regions near the conformal boundaries are most easily investigated 
in the AdS-Schwarzschild charts (\ref{eq:SchwmetricDimensionful}). Letting 
$z = \epsilon$ and taking the limit $\epsilon \to 0$
the other coordinates behave as 
\eqn{ \label{eq:epsilons}
  w_+ &=& \pm (1-\epsilon/z_H) e^{t/z_H} + \cO(\epsilon^2),  \nn \\
  w_- &=& \pm (-1+\epsilon/z_H) e^{-t/z_H} + \cO(\epsilon^2), \nn \\
  1 + w_+ w_- &=& 2\epsilon/z_H + \cO(\epsilon^2), \nn \\
  1 - w_+ w_- &=&  2-2\epsilon/z_H + \cO(\epsilon^2),
}
where the $\pm$ sign is chosen in region $R_\pm$.

\subsubsection{Embedding distance}
\label{sec:distance}

The embedding space endows the BTZ geometry with a convenient notion
of distance. We define the \emph{embedding distance} between two points
on BTZ to be $\Theta(X,Y) := g_{AB}X^A Y^B / \ell^2$. The embedding
distance is simply related to the length of the chord through the 
embedding space
\eq{
  ||X - Y||^2 = -2 \ell^2 (\Theta(X,Y) + 1) ,
}
as well as the geodesic distance $D(X,Y)$ on the BTZ manifold:
\eq{
  D(X,Y) = 
  \left\{ 
    \begin{array}{ll}
      \ell \,\text{acosh}\left[-\Theta(X,Y)\right]  
      & \quad \text{spacelike separation} \\
      \ell \,\text{acos}\left[-\Theta(X,Y)\right] 
      & \quad \text{timelike separation}.
    \end{array}
  \right.
}
In the coordinates (\ref{eq:wmetric}) $\Theta(X,Y)$ may be written
\eq{ \label{eq:Theta}
  \Theta(X_1, X_2) = 
 - \frac{2 (w_{+1} w_{-2} + w_{-1} w_{+2}) 
    +(1-w_{+1} w_{-1})(1-w_{+2} w_{-2})\cosh(\phi_1-\phi_2)}
  {(1+w_{+1} w_{-1})(1+w_{+2} w_{-2})} .
}
Both the embedding distance and geodesic distance generically diverge 
as one or both points approach the conformal boundaries. Using the limits
(\ref{eq:epsilons}) we compute the limit of the embedding distance
when both points approach a boundary:
\eqn{ \label{eq:Thetalimit}
  \Theta(X_1,X_2) 
  &=& \frac{\Theta^{\rm reg}(X_1,X_2)}{\epsilon_1 \epsilon_2}
  + \text{finite} , \nn \\
  \Theta^{\rm reg}(X_1,X_2) 
  &:=& z_H^2 \left[
 s(X_1,X_2)  \cosh\left(\frac{t_1 - t_2}{z_H}\right) - \cosh\left( \frac{\theta_1-\theta_2}{z_H}\right) \right],
}
where $s(X_1,X_2) = +(-)$ for points approaching the same 
(different) boundaries. 
For points which are spacelike separated in this limit the geodesic 
distance behaves as
\eq{ \label{eq:Dlimit}
  \ell^{-1} D(X_1,X_2) = -\ln(\epsilon_1 \epsilon_2) 
  + \ln\left[ - 2 \Theta^{\rm reg}(X_1,X_2)\right] + \dots,
  \quad X_1, X_2\;\text{spacelike},
}
with ellipses denoting terms that vanish as $\epsilon_{1,2} \to 0$.
In examining these relations it is useful to recall that the 
AdS-Schwarzschild coordinate $t$ flows in opposite directions on the 
two boundaries. For instance, under the action of a boost with positive 
rapidity both $t_1$ and $t_2$ increase, regardless of which region 
the points belong, while the difference $t_1-t_2$ is unaffected.

\subsection{Holographic computation}
\label{sec:computation}

The embedding space description makes it quite
easy to compute both the holographic mutual information
and holographic thermo-mutual information.
Both quantities may be computed from the dimensionless ``cross ratio''
\eqn{ \label{eq:J}
  J &:=& \ell^{-1} \left[D(X_{A1},X_{A2}) + D(X_{B1},X_{B2}) 
    - D(X_{A1},X_{B1}) - D(X_{A2},X_{B2}) \right] ,\quad
}
which we define for four points $X_{A1}$, $X_{A2}$, $X_{B1}$, $X_{B2}$ 
in $R_+ \cup R_-$. To obtain the HMI let the four points be in
$R_+$ with AdS-Schwarzschild coordinates $X = (t,z,\theta)$:
\eqn{ \label{eq:Xs}
  X_{A1} = (0,\epsilon,\theta_{A1}) , \quad
  X_{A2} = (0,\epsilon,\theta_{A2}) , \quad
  X_{B1} = (0,\epsilon,\theta_{B1}) , \quad
  X_{B2} = (0,\epsilon,\theta_{B2}) ,
}
and
\eq{ \label{eq:thetas}
  0 \le \theta_{A1} < \theta_{A2} < \theta_{B2} < \theta_{B1} < 2\pi .
}
The HMI is then given by
\eq{
  \HMI(A_1 : B_1) := \frac{\ell}{4 G_N} \max \left[
    \lim_{\epsilon \to 0} J, \; 0 \right] .
}
Taking the limit with the aid of (\ref{eq:Dlimit}) and 
(\ref{eq:Thetalimit}) we obtain
\eqn{ \label{eq:HMI}
  \HMI(A_1 : B_1) 
  &=& 
  \frac{\ell}{4 G_N} \ln {\rm max}
  \left[
    \frac{\left(1 - \cosh\frac{\Delta\theta_A}{z_H}\right)
      \left(1 - \cosh\frac{\Delta \theta_B}{z_H}\right)}
    {\left(1 - \cosh\frac{\Delta\theta_1}{z_H}\right)
      \left(1 - \cosh\frac{\Delta\theta_2}{z_H}\right)} 
    ,\; 1 \right] ,
}
where we have introduced the obvious notation
\eq{
  \Delta \theta_A = |\theta_{A1}-\theta_{A2}| , \quad
  \Delta \theta_B = |\theta_{B1}-\theta_{B2}| , \quad
  \Delta \theta_1 = |\theta_{A1}-\theta_{B1}| , \quad
  \Delta \theta_2 = |\theta_{A2}-\theta_{B2}| .
}
The ordering (\ref{eq:thetas}) is a convenient choice
of labelling so that (\ref{eq:HMI}) is constructed from
the correct choice of geodesics such that the surface $M_C$
is homologous to $A_1 \cup B_1$.

The HTMI is similarly obtained from (\ref{eq:J}) by considering,
e.g., $X_{A1},X_{A2}\in R_+$ and $X_{B1},X_{B2}\in R_-$.
In their respective AdS-Schwarzschild charts
the points may once again be labelled by
(\ref{eq:Xs}) with angles satisfying (\ref{eq:thetas}). 
This choice of labelling selects the correct surface $M_C$ 
for defining $S_{A1 \cup B2}$ -- recall discussion in \S\ref{sec:HTMI}.
The HTMI is then
\eq{ 
  \HTMI(A_1 : B_2) :=
   \frac{\ell}{4 G_N} \max \left[ 
     \lim_{\epsilon \to 0} J(A : B), \; 0 \right] ,
}
and taking the limit we obtain
\eq{ \label{eq:HTMI}
  \HTMI(A_1 : B_2) 
    = \frac{\ell}{4 G_N} \ln {\rm max}
    \left[
    \frac{\left(1 - \cosh\frac{\Delta\theta_A}{z_H}\right)
      \left(1 - \cosh\frac{\Delta\theta_B}{z_H}\right)}
    {\left(1 + \cosh\frac{\Delta\theta_1}{z_H}\right)
      \left(1 + \cosh\frac{\Delta\theta_2}{z_H}\right)} 
      ,\; 1 \right] . 
}

\begin{figure}[h!]
  \centering
  \includegraphics[width=0.5\textwidth]{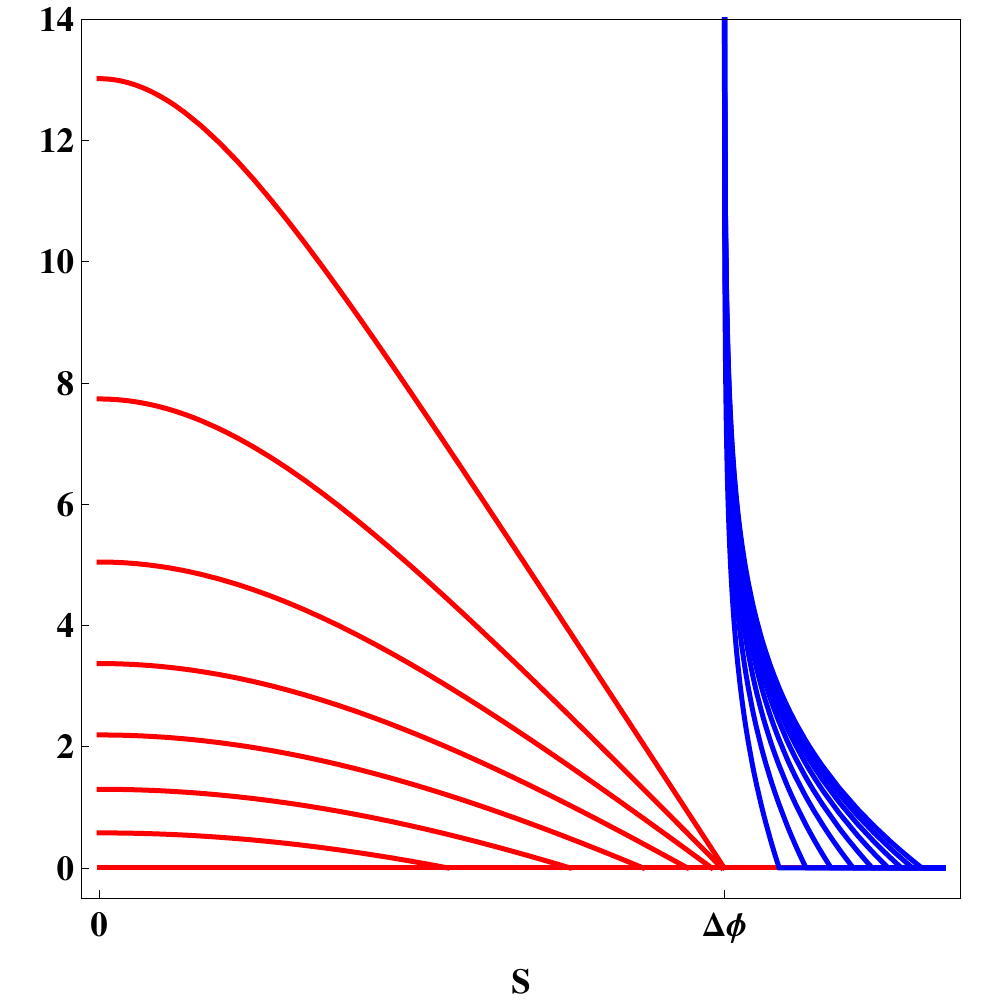}
  \caption{An example of the holographic mutual information (HMI) 
    and thermo-mutual information (HTMI). For this plot the angular 
    widths are $\Delta\phi_A = \Delta\phi_B = \Delta\phi$; $S$ denotes the
    angular separation between the centers of $A$ and $B$ respectively.
    In blue are profiles of the HMI for various temperatures, with 
    increasing temperature yielding a non-zero HMI at larger $S$.
    In red are profiles of the HTMI for the same temperatures,
    with increasing temperature yielding greater HTMI. Note that
    the HTMI vanishes when the $\theta$ profiles of $A$ and $B$ have 
    no overlap. For sufficiently low temperature,
    though still above the Hawking-Page transition, the HTMI vanishes
    for all $S$. These features are generic.
    \label{fig:holographic}}
\end{figure}

As with the thermo-mutual information computed in the quantum
mechanics and CFT examples above, the HTMI captures the basic properties
of thermo-mutual information outlined in \S\ref{sec:TMI}, though
the agreement is not perfect. In particular we note:
\begin{enumerate}[i)]
\item By construction HTMI is non-negative.
\item From (\ref{eq:HTMI}) it is clear that HTMI is bounded above for 
  all configurations. The fact that HTMI is constructed from two regions
  on different boundaries makes it manifest that no UV divergences
  are encountered when $A_1$ and $B_2$ are taken to have overlapping 
  $\theta$ profiles. 
\item Given two regions $B_{1,2}$ with equivalent $\theta$
  profiles, one region on each boundary, we obtain from
  (\ref{eq:HTMI}) and (\ref{eq:HMI}) that
  $\HMI(A_1 : B_1) - \HTMI(A_1 : B_2) \ge  0$. In this sense HTMI bounds
  HMI from below.
\item
    The high temperature limit:
    when $\beta$ is much less than all length scales $\Delta \theta_A$,
    $\Delta \theta_B$, $\Delta \theta_1$, $\Delta \theta_2$,
    the HTMI becomes
    \eq{
      \HTMI(A_1 : B_1) \approx \frac{2\pi \ell T}{4 G_N}
      \max\left[ (\Delta \theta_A + \Delta \theta_B
        - \Delta \theta_1 - \Delta \theta_2), 0 \right] ,
    }
    where we have reinstated $T$ for clarity. This expression is non-vanishing
    when $A_1$ and $B_2$ have overlapping $\theta$ profile, and
    in this case $\HTMI(A_1 : B_2) \approx 2 s (L/\ell)$, where $s$ is the entropy density and $L/\ell$ is the
    length of the overlapping region in AdS units. We conjecture that this behavior is generic and that for any equilibrium finite temperature system, in the high temperature limit $TMI \approx 2 s \mathrm{Vol(overlap)}$.
\item
Behavior at low temperature:
  we may readily verify from (\ref{eq:HTMI}) that 
  $\HTMI(A_1 : B_2)$ tends to zero in the limit $\beta \to \infty$.
  In fact, the HTMI vanishes for $\beta \sim \pi \sqrt{\theta_A \theta_B}$,
  which can occur well above the Hawking-Page temperature $\beta = 2\pi$. We conjecture that this will occur generically even in higher dimensions.
    
\end{enumerate}
This last point, the fact that the HTMI can vanish at temperatures 
well above the Hawking-Page temperature, deserves further discussion.
In \S\ref{sec:TMI} we made a point to note that at any finite temperature
the TMI cannot vanish for all configurations. This follows from the
fact that the TMI bounds the correlations of type-1 and type-2 fields, 
and these correlations cannot everywhere vanish if $\Omega$ is indeed
entangled. The fact that the HTMI can vanish for typical regions at 
finite temperature is a large $N$ artifact. Recall that the RT formula is believed to 
capture only the $\cO(N^2)$ contributions to the entanglement entropies. 
Therefore the HTMI is itself contains only the leading large $N$ behavior
of the TMI. Correlations between type-1 and type-2 fields are naturally 
$\cO(N^0)$ are therefore are technically zero at this order. 
We provide a representative plot of HTMI and HMI in Fig.~\ref{fig:holographic}.

\section{Discussion}
\label{sec:discussion}

In this paper we have studied thermo-mutual information (TMI), an analogue
of mutual information which provides a measure for the correlation
between `physical' and thermo-double degrees of freedom.
The basic attributes of TMI may be determined directly from its 
definition in thermal QFT, are quite analogous to those of mutual
information, and are summarized in \S\ref{sec:TMI}. 
We highlight in particular that TMI is a UV-finite quantity in field 
theory  is ``universal'' (i.e. UV regulator-independent) and is well-defined for any theory local enough to factor the Hilbert space or operators.
Through the Pinsker inequality (\ref{eq:TMIPinsker})
TMI provides an upper bound on the correlation functions between
type-1 and type-2 operators inserted at arbitrary separations.
As a result, TMI carries much more qualitative information about
a state of a theory than the entropy of the mixed state and has different information than the mutual information.
We have explored TMI in detail in three examples: a two-site spin chain,  a 2d Dirac fermion, and a 2d CFT described by a holographic dual.

It is natural to ask what the possible applications of TMI are within
thermal field theory. As a simple calculation, consider the computation
of correlators in the thermal vacuum using canonical
perturbation theory about small couplings. In this setting the
correlation function $\C{\phi_1(x) \phi_2(y)}_\Omega$ provides one
of the Schwinger-Keldysh Green's functions. TMI may be used to 
estimate the strength of perturbative corrections; in particular,
TMI may be used to bound from above the smooth, UV-finite part of
Feynman diagrams arising from the coupling between type-1 and type-2
fields. Similarly, TMI may be used to bound the smooth part of
retarded and advanced thermal Green's functions.

Of course, our primary motivation for introducing TMI was so
that we could study it holographically via AdS/CFT. 
There are several reasons for this. First, we note that holographic
TMI provides yet another test of the conjectured Ryu-Takayanagi 
formula \cite{Ryu:2006bv}. Indeed, as described in \S\ref{sec:TMI},
holographic TMI has all the hallmarks of a TMI in (large $N$) field theory.
Certainly more interesting is the fact that holographically-computed
TMI provides the simplest example of holographic mutual information 
between regions on disconnected AdS boundaries.
In general, the dual CFT description of AdS spaces with multiple boundaries is 
poorly understood. It is reasonable to expect that holographic mutual 
information can provide basic information about the CFT which describes 
these bulk spacetimes. Perhaps the most obvious example is to note
that holographic mutual information, if it indeed provides a measure
of mutual information in the dual CFT, bounds the correlations
between operators on different boundaries.

In general the agreement between the holographically-computed
TMI and our expectations from QFT is quite good, but we should 
comment on the most significant discrepancy: the holographic TMI can 
vanish for any choice of region at finite temperature, contrary the
definition of TMI. This is due to the fact that the holographic
computation captures only the leading behavior in $N^2$, and so like
the holographic mutual information exhibits features similar to a first
order phase transition. Indeed, that it is precisely the same 
phenomena as for the holographic mutual information may be seen as follows.
One may adopt Rindler coordinates on the boundary of global
AdS and in doing so describe the zero-temperature pure state (from
the global perspective) as a thermal state at the Rindler temperature.
The CFT in one Rindler wedge may be interpreted in the TFD language as
the thermo-double of the the CFT in the other wedge. 
In this set-up the thermo-mutual information is precisely the mutual information 
of the global perspective.

We close with a discussion of some possible future directions of research.

\subsection{Holographic TMI in higher dimensions}

A natural extension of this work would be to examine holographic TMI 
computed in higher dimensional black hole backgrounds. While conceptually 
a straight-forward generalization, the necessary computations are considerably
more involved in $d+1$ bulk dimensions with $d>2$. Partly this is due
to the fact that in higher dimensions the covariant form of the RT formula 
is necessary and the extremal surfaces involved have dimension greater than
two. In addition, in $2+1$ bulk dimensions our computations were rather 
slick because of the fact that BTZ is a quotient of $AdS_3$.
Regardless of these computational challenges, we expect similar physics
in higher dimensions. Namely, we expect the holographically computed 
TMI to be non-vanishing even at leading order in $N^2$, and we expect it to 
exhibit a first order phase transition at $\beta \sim \pi L$ where $L$ is 
the length scale of the regions under consideration.
The features of TMI listed in \S\ref{sec:TMI} are of course independent
of dimension.

\subsection{Other asymptotically-AdS geometries}

It has been pointed out by several authors \cite{Aminneborg:1997pz,
  Brill:1998pr,Krasnov:2000zq,Krasnov:2003aa,Skenderis:2009ju}
that in $2+1$ Einstein gravity the maximally extended BTZ geometry is 
not the only solution which has one asymptotic region corresponding to
that of the AdS-Schwarzschild chart \eqref{eq:SchwmetricDimensionful}. 
There also exist solutions with more than two asymptotic regions
as well as those with only one asymptotic region but non-trivial topology
behind the horizon.  We expect holographically-computed mutual information between regions
  on separate boundaries to be useful in characterizing the CFT
  dual description of these spacetimes.
Geometries with more than two asymptotic regions may be thought of 
as describing different choices of purification for a system initially
described by a thermal density matrix. These purifications enlarge the 
total Hilbert space to a space other than the TFD Hilbert space 
$\cH^{\otimes2}$. Basic features of these purifications may 
be determined by examining $\HTMI$-like mutual information ``correlators.''

Solutions with only one asymptotic region but non-trivial topology
are even more fascinating. The AdS/CFT intuition states that the CFT
dual of these geometries must be in a pure state. However, the RT 
formula applied in any black hole spacetime predicts a mixed state:
due to the presence of the black hole in the bulk, the RT formula computes
for any region $A$ that $S_A \neq S_{\bar{A}}$, indicating that the state
is mixed. For these spacetimes it appears that the RT formula gives
the incorrect result at leading order in $N^2$, but this warrants
a more careful analysis.

\subsection{Renormalized entanglement entropy and mutual information}

We have emphasized throughout our analysis that one important reason mutual information (and, for intersecting regions, the $F$-function) is an object of 
interest is that it is a manifestly UV finite quantity. 
Recently \cite{Liu:2012eea} has proposed a UV-finite ``renormalized 
entanglement entropy'' (REE). It is interesting to ask if the same physics
contained the MI may also be recovered from linear combinations of the
REE. For a 2d system \cite{Liu:2012eea} define the REE
for a subregion $A$ with length scale $L$ to be $\cS_A = L \partial_L S_A$. 
The naive ``renormalized mutual information'' is then
$L \partial_L (S_A + S_B - S_{A \cup B})$. However, this quantity does
not appear to capture the same physics; in particular, the derivative
with respect to $L$ annihilates the dependence on the 
scale-invariant cross-ratio. For zero temperature CFTs
the mutual information is a function only of this cross-ratio,
so this renormalized mutual information vanishes. For CFTs at finite 
temperature the naive renormalized mutual information is non-zero but
again does not depend on the cross-ratio and so does not contain
information about correlations, only information about the geometries
under consideration. We mention this here in the hope that it spurs a
better understanding of the relationship between the proposed REE
and the UV-finite quantities considered in our analysis.\\

\noindent{\bf Acknowledgments}\\

We thank Matthew Headrick, Matthew Kleban, Alex Maloney, Guy Moore, and
Massimo Porrati for useful discussions. MMR would like to thank Hurricane Sandy for a stimulating work environment while this project was being completed. IAM and MMR are supported 
by the Simons Postdoctoral Fellowship Program.

\appendix 

\section{Euclidean path integrals for reduced density matrices
in TFD}
\label{app:PI}

In this appendix we derive the results quoted in \S\ref{sec:QFT}.
We begin by reviewing the relationship between the TFD path integral
and the Euclidean path integral.
Boundary conditions must be supplied in order for
any path integral expression to be well-defined;
for thermal systems the relevant boundary condition is the 
KMS condition \cite{Haag:1967aa}.
Denoting the time translation of $\phi$ by
$\phi(t) = e^{i H t} \phi(0) e^{-i H t}$ and suppressing other 
spacetime arguments, one readily deduces from (\ref{eq:Omega1})
that correlators of the thermal state $\Omega$ at inverse temperature
$\beta$ satisfy
\eq{ \label{eq:KMS}
  \C{ \phi(t) \chi }_\Omega 
  = \C{ \chi \phi(t + i\beta) }_\Omega ,
}
for two operators $\phi(x)$ and $\chi(x)$. 
A consequence of (\ref{eq:KMS}) is that
correlators with respect to $\Omega$ are boundary values of
analytic functions in the complex time domain $\Im t \in (0,-\beta)$
satisfying the condition (\ref{eq:KMS}). Therefore we may
seek a path integral generating function $Z_C[J]$ 
for the path-ordered correlation functions
$\C{ P \phi(x_1)\dots\phi(x_n) }_\Omega$. The `path' refers to
the time integration which is traversed from $\Im t = 0$ to
$\Im t = -\beta$ along a contour $C$; a more careful examination of the analyticity
properties of such correlators reveals that the path must have
monotonically non-increasing $\Im t$ \cite{Landsman:1986uw}.
The path integral may be constructed in the usual manner:
\eq{ \label{eq:ZC}
  Z_C[J] := Z^{-1} \int [\cD\phi]\, \exp\left[
    iS_C[\phi]
    +i\int_C dt \int d^{D-1}x \sqrt{-g(x)} J(x) \phi(x)
  \right] ,
}
$D$ is the number of spacetime dimensions and $Z = Z[0]$. 
The subscript $S_C[\phi]$ denotes that, like
the source term written explicitly in (\ref{eq:ZC}), the time
integration in the action is over the path $C$.
The boundary conditions imposed on the field integration is
$\phi(t) = \phi(t+i\beta)$. Path-ordered correlation functions
may be computed by taking functional derivatives with respect
to the source $J(x)$:
\eq{
  \C{ P \phi(x_1)\dots\phi(x_n) }_\Omega
  =  \frac{\delta^n Z_C[J]}{i\delta J(x_1)\cdots i \delta J(x_n)}
  \bigg|_{J=0}  .
}

The two choices of time contour relevant for our discussion 
are shown in Fig.~\ref{fig:SKcontour}. The first choice is the
Euclidean contour which is traversed along the imaginary $t$ axis 
$0 \to -i\beta$. This path integral determines the
Euclidean correlators $\C{ \phi(x_1)\dots\phi(x_n) }_{\rm E}$
which define the Euclidean `state.' Explicitly,
\eq{ \label{eq:ZE}
  Z_{\rm E}[J] := Z^{-1} \int [\cD\phi]\, \exp\left[
    -S_{\rm E}[\phi]
    -\int_{-\beta}^0 d\tau \int d^{D-1}x \sqrt{g(x)} J(x) \phi(x)
  \right] .
}
A second time contour is the TFD contour -- see Fig.~\ref{fig:SKcontour}.
With $t_i$ and $t_f$ real initial and final times respectively, this 
contour is traversed: i) $t_i \to t_f$, ii) $t_f \to t_f-i\beta/2$,
iii) $t_f -i\beta/2 \to t_i -i\beta/2$, iv) $t_i-i\beta/2 \to t_i -i \beta$.
Under reasonable circumstances (see discussion 
\S2.4 of \cite{Landsman:1986uw}) the path integral factorizes into two
path integrals, one involving the horizontal legs of the contour, the
other involving the vertical legs. If one is interested only in 
correlators of operators lying on the horizontal contours then the
latter path integral is simply $1$ due to the overall normalization.
The path integral for the horizontal segments is
\eqn{ \label{eq:ZTFD}
  Z_{\rm TFD}[J_1,J_2] &:=& 
  Z_{\rm TFD}^{-1} \int [\cD\phi_1][\cD\phi_2]\, \exp\Bigg[
    +iS[\phi_1] - iS[\phi_2]
    \nn \\ & &
    \phantom{\cN \int }
    +i \int d^Dx \sqrt{g(x)} \left(J_1(x) \phi_1(x) - J_2(x) \phi_2(x)\right)
  \Bigg] .
}
Here $\phi_1(x)$ denote fields on the real axis while $\phi_2(x)$ fields
have $\Im t = -\beta/2$. As our notation implies, these fields correspond
to the type-1 and type-2 operators described in \S\ref{sec:thermal}.
Correlators of the operators are generated via:
\eqn{
  & & \C{ \left[T \phi_1(x_1)\dots\phi_1(x_n)\right]
    \left[\overline{T} \phi_2(y_1)\dots \phi_2(y_m)\right] }_\Omega
  \nn \\ & &
  \quad\quad\quad\quad =  \frac{\delta^{n+m} Z_{\rm TFD}[J_1,J_2]}
  {i\delta J_(x_1)\cdots i \delta J_1(x_n)
    (-i)\delta J_2(y_1)\cdots (-i) \delta J_2(y_m)}
  \bigg|_{J_1=J_2=0} .
}

We may now construct path integral representations
of a reduced density matrix in the TFD system.
Let us first consider the density operator
$\rho_{A1} = \rho(\Omega; \cA_1(A))$. Without loss of generality
we let the region $A \subset \Sigma_0$.
For a basis in $\cH_{A1}$ we use the set of states $\{\psi_{A1}\}$
specified by their field configuration of type-1 fields $\psi_1(x)$ on 
$A$:
\eq{
  \psi_{A1} := \{\phi_1(x) = \psi_1(x) \;|\; x\in A\} .
}
We denote the matrix elements of $\rho_{A1}$ in this basis
$\rho_{A1}(\psi'_{A1}|\psi_{A1})$.
Following \cite{Calabrese:2004eu}, the path integral representation
of $\rho_{A1}(\psi'_{A1}|\psi_{A1})$ is then easily constructed by inserting
delta functionals into the path integral (\ref{eq:ZTFD}):
\eqn{ \label{eq:rhoA1}
  \rho_{A1}(\psi'_{A1}|\psi_{A1}) &=&
  Z^{-1} \int [\cD\phi_1][\cD\phi_2] \Bigg\{ \exp\left[
    +iS[\phi_1] - iS[\phi_2]\right]
  \nn \\ & & 
  \left[\prod_{x\in A} 
  \delta(\phi_1(0^+)-\psi'_1(0^+))
  \delta(\phi_1(0^-)-\psi_1(0^-))
  \right]\Bigg\} .
}
We have suppressed the dependence of spatial coordinates.
The limits $0^\pm$ denote how $t$ should be taken to approach 
$t=0$, and reflect the fact that we define $\psi'$ ($\psi$) in the
limit $t \to 0^{+(-)}$.
The analogous expression for the density matrix
$\rho_{A2}$, in an analogous basis $\{\psi_{A2}\}$, is
\eqn{ \label{eq:rhoA2}
  \rho_{A2}(\psi'_{A2}|\psi_{A2}) &=&
  Z^{-1} \int [\cD\phi_1][\cD\phi_2] \Bigg\{ \exp\left[
    +iS[\phi_1] - iS[\phi_2]\right]
  \nn \\ & & 
  \Bigg[\prod_{x\in A} 
  \delta(\phi_2(-i\beta/2+0^+)-\psi'_2(-i\beta/2+0^+))
  \nn \\ & & \phantom{  \Bigg[\prod_{x\in A} }
  \delta(\phi_2(-i\beta/2+0^-)-\psi_2(-i\beta/2+0^-))
  \Bigg]\Bigg\} .
}
The density matrices $\rho_{A1\cup B1}$ and $\rho_{A1 \cup B2}$ are
constructed from the obvious generalizations of these expressions.

We would like to compute these path integrals in Euclidean signature;
this requires that we deform the TFD time contour to the Euclidean contour.
To do so we recast the $t+0^\pm$ limits in the expressions above
in terms of path ordering rather than real time ordering.
Let $\lambda(t)$ denote a function that decreases monotonically along the 
path from $t_i$ to $t_i-i\beta$. 
By rewriting the TFD path integral in terms of $\lambda(t)$ it
is then straightforward to deform the time contour to the Euclidean
contour. We then let $\lambda = \Im t$.
For instance, after performing this step we obtain a Euclidean
expression for the density matrix $\rho_{A1}$:
\eqn{ \label{eq:rhoA1E}
  \rho_{A1}^{\rm E}(\psi'_A|\psi_A) 
  &=& Z^{-1} \int [\cD\phi]\, e^{-S_{\rm E}[\phi]}
  \Bigg[\prod_{x\in A} 
  \delta(\phi(0^-)-\psi_1'(0^-))
  \delta(\phi(0^+)-\psi_1(0^+))
  \Bigg] ,\nn \\
}
where now we label only the dependence on $\Im t$.
By the same steps we obtain the analogue of $\rho_{A2}$:
\eqn{ \label{eq:rhoA2E}
  \rho_{A2}^{\rm E}(\psi'_A|\psi_A) 
  &=& Z^{-1} \int [\cD\phi]\, e^{-S_{\rm E}[\phi]}
  \Bigg[\prod_{x\in A} 
  \delta(\phi(-\beta/2+0^+)-\psi_2'(-\beta/2+0^+))
  \nn \\ & & \phantom{Z^{-1} \int [\cD\phi]\, e^{-S_{\rm E}[\phi]}\Bigg[\,}
  \delta(\phi(-\beta/2+0^-)-\psi_2(-\beta/2+0^-))
  \Bigg] .\nn \\
}
Notice that the $i\epsilon$ prescriptions defining the states
have switched.
Finally, we record the formula for the Euclidean analogue
of the density matrix $\rho_{A1 \cup B2}$. This is the key formula
necessary to compute the thermo-mutual information via the
replica trick:
\eqn{ \label{eq:rhoABE}
  & &\rho_{A1 \cup B2}^{\rm E}(\psi'|\psi) 
  \nn \\
  &=& Z^{-1} \int [\cD\phi]\, e^{-S_{\rm E}[\phi]} \Bigg\{
  \Bigg[\prod_{x\in A} 
  \delta(\phi(0^-)-\psi'(0^-))
  \delta(\phi(0^+)-\psi(0^+))
  \Bigg] 
  \nn \\ & & \phantom{Z^{-1} \int [\cD\phi]\, e^{-S_{\rm E}[\phi]}\Bigg\{}
  \Bigg[\prod_{x\in B} 
  \delta(\phi(-\beta/2+0^+)-\psi'(-\beta/2+0^+))
  \nn \\ & & \phantom{Z^{-1} \int [\cD\phi]\, e^{-S_{\rm E}[\phi]}\Bigg\{
  \prod{blah} \;}
  \delta(\phi(-\beta/2+0^-)-\psi(-\beta/2+0^-))
  \Bigg] \Bigg\} . 
}


\addcontentsline{toc}{section}{Bibliography}
\bibliographystyle{JHEP}
\bibliography{TMI_refs}

\providecommand{\href}[2]{#2}\begingroup\raggedright\begin{thebibliography}{10}

\bibitem{Groisman:2005aa}
B.~Groisman, S.~Popescu, and A.~Winter, {\it Quantum, classical, and total
  amount of correlations in a quantum state},  {\em Phys. Rev. A} {\bf 72}
  (Sep, 2005) 032317, [\href{http://xxx.lanl.gov/abs/quant-ph/0410091}{{\tt
  quant-ph/0410091}}].

\bibitem{Wolf:2008aa}
M.~M. Wolf, F.~Verstraete, M.~B. Hastings, and J.~I. Cirac, {\it Area laws in
  quantum systems: Mutual information and correlations},  {\em Phys. Rev.
  Lett.} {\bf 100} (Feb, 2008) 070502,
  [\href{http://xxx.lanl.gov/abs/0704.3906}{{\tt arXiv:0704.3906}}].

\bibitem{Calabrese:2004eu}
P.~Calabrese and J.~L. Cardy, {\it {Entanglement entropy and quantum field
  theory}},  {\em J.Stat.Mech.} {\bf 0406} (2004) P06002,
  [\href{http://xxx.lanl.gov/abs/hep-th/0405152}{{\tt hep-th/0405152}}].

\bibitem{Calabrese:2009qy}
P.~Calabrese and J.~Cardy, {\it {Entanglement entropy and conformal field
  theory}},  {\em J.Phys.A} {\bf A42} (2009) 504005,
  [\href{http://xxx.lanl.gov/abs/0905.4013}{{\tt arXiv:0905.4013}}].

\bibitem{Casini:2009sr}
H.~Casini and M.~Huerta, {\it {Entanglement entropy in free quantum field
  theory}},  {\em J.Phys.A} {\bf A42} (2009) 504007,
  [\href{http://xxx.lanl.gov/abs/0905.2562}{{\tt arXiv:0905.2562}}].

\bibitem{Ryu:2006bv}
S.~Ryu and T.~Takayanagi, {\it {Holographic derivation of entanglement entropy
  from AdS/CFT}},  {\em Phys. Rev. Lett.} {\bf 96} (2006) 181602,
  [\href{http://xxx.lanl.gov/abs/hep-th/0603001}{{\tt hep-th/0603001}}].

\bibitem{Michalogiorgakis:2008aa}
G.~Michalogiorgakis, {\it Entanglement entropy of two dimensional systems and
  holography},  {\em Journal of High Energy Physics} {\bf 2008} (2008), no.~12
  068, [\href{http://xxx.lanl.gov/abs/0806.2661}{{\tt arXiv:0806.2661}}].

\bibitem{Schwimmer:2008aa}
A.~Schwimmer and S.~Theisen, {\it Entanglement entropy, trace anomalies and
  holography},  {\em Nuclear Physics B} {\bf 801} (2008), no.~1--2 1 -- 24,
  [\href{http://xxx.lanl.gov/abs/0802.1017}{{\tt arXiv:0802.1017}}].

\bibitem{Headrick:2010zt}
M.~Headrick, {\it {Entanglement Renyi entropies in holographic theories}},
  {\em Phys.Rev.} {\bf D82} (2010) 126010,
  [\href{http://xxx.lanl.gov/abs/1006.0047}{{\tt arXiv:1006.0047}}].

\bibitem{Casini:2011kv}
H.~Casini, M.~Huerta, and R.~C. Myers, {\it {Towards a derivation of
  holographic entanglement entropy}},  {\em JHEP} {\bf 1105} (2011) 036,
  [\href{http://xxx.lanl.gov/abs/1102.0440}{{\tt arXiv:1102.0440}}].

\bibitem{Hubeny:2007xt}
V.~E. Hubeny, M.~Rangamani, and T.~Takayanagi, {\it {A Covariant holographic
  entanglement entropy proposal}},  {\em JHEP} {\bf 0707} (2007) 062,
  [\href{http://xxx.lanl.gov/abs/0705.0016}{{\tt arXiv:0705.0016}}].

\bibitem{Hubeny:2012ry}
V.~E. Hubeny, {\it {Extremal surfaces as bulk probes in AdS/CFT}},  {\em JHEP}
  {\bf 1207} (2012) 093, [\href{http://xxx.lanl.gov/abs/1203.1044}{{\tt
  arXiv:1203.1044}}].

\bibitem{Hubeny:2012wa}
V.~E. Hubeny and M.~Rangamani, {\it {Causal Holographic Information}},  {\em
  JHEP} {\bf 1206} (2012) 114, [\href{http://xxx.lanl.gov/abs/1204.1698}{{\tt
  arXiv:1204.1698}}].

\bibitem{Czech:2012bh}
B.~Czech, J.~L. Karczmarek, F.~Nogueira, and M.~Van~Raamsdonk, {\it {The
  Gravity Dual of a Density Matrix}},  {\em Class.Quant.Grav.} {\bf 29} (2012)
  155009, [\href{http://xxx.lanl.gov/abs/1204.1330}{{\tt arXiv:1204.1330}}].

\bibitem{Bousso:2012sj}
R.~Bousso, S.~Leichenauer, and V.~Rosenhaus, {\it {Light-sheets and AdS/CFT}},
  {\em Phys.Rev.} {\bf D86} (2012) 046009,
  [\href{http://xxx.lanl.gov/abs/1203.6619}{{\tt arXiv:1203.6619}}].

\bibitem{Marolf:2012xe}
D.~Marolf and A.~C. Wall, {\it {Eternal Black Holes and Superselection in
  AdS/CFT}},  \href{http://xxx.lanl.gov/abs/1210.3590}{{\tt arXiv:1210.3590}}.

\bibitem{Maldacena:2001kr}
J.~M. Maldacena, {\it {Eternal black holes in anti-de Sitter}},  {\em JHEP}
  {\bf 0304} (2003) 021, [\href{http://xxx.lanl.gov/abs/hep-th/0106112}{{\tt
  hep-th/0106112}}].

\bibitem{Casini:2003ix}
H.~Casini, {\it Geometric entropy, area and strong subadditivity},  {\em
  Classical and Quantum Gravity} {\bf 21} (2004), no.~9 2351,
  [\href{http://xxx.lanl.gov/abs/hep-th/0312238}{{\tt hep-th/0312238}}].

\bibitem{Casini:2004bw}
H.~Casini and M.~Huerta, {\it {A Finite entanglement entropy and the
  c-theorem}},  {\em Phys.Lett.} {\bf B600} (2004) 142--150,
  [\href{http://xxx.lanl.gov/abs/hep-th/0405111}{{\tt hep-th/0405111}}].

\bibitem{Eisert:2010fk}
J.~Eisert, {\it {Area laws for the entanglement entropy}},  {\em Reviews of
  Modern Physics} {\bf 82} (2010), no.~1 277--306,
  [\href{http://xxx.lanl.gov/abs/0808.3773}{{\tt arXiv:0808.3773}}].

\bibitem{Refael:2009aa}
G.~Refael and J.~E. Moore, {\it Criticality and entanglement in random quantum
  systems},  {\em Journal of Physics A: Mathematical and Theoretical} {\bf 42}
  (2009), no.~50 504010, [\href{http://xxx.lanl.gov/abs/0908.1986}{{\tt
  arXiv:0908.1986}}].

\bibitem{Schwinger:1960qe}
J.~S. Schwinger, {\it {Brownian motion of a quantum oscillator}},  {\em
  J.Math.Phys.} {\bf 2} (1961) 407--432.

\bibitem{Keldysh:1964ud}
L.~V. Keldysh, {\it {Diagram technique for nonequilibrium processes}},  {\em
  Zh. Eksp. Teor. Fiz.} {\bf 47} (1964) 1515--1527.

\bibitem{Landsman:1986uw}
N.~P. Landsman and C.~G. van Weert, {\it {Real and Imaginary Time Field Theory
  at Finite Temperature and Density}},  {\em Phys. Rept.} {\bf 145} (1987) 141.

\bibitem{Takahashi:1996zn}
Y.~Takahashi and H.~Umezawa, {\it {Thermo field dynamics}},  {\em
  Int.J.Mod.Phys.} {\bf B10} (1996) 1755--1805.

\bibitem{Sakurai:1993aa}
J.~J. Sakurai, {\em {Modern Quantum Mechanics (Revised Edition)}}.
\newblock Addison Wesley, 1~ed., Sept., 1993.

\bibitem{Calabrese:2012nk}
P.~Calabrese, J.~Cardy, and E.~Tonni, {\it {Entanglement negativity in extended
  systems: A field theoretical approach}},
  \href{http://xxx.lanl.gov/abs/1210.5359}{{\tt arXiv:1210.5359}}.

\bibitem{Calabrese:2012ew}
P.~Calabrese, J.~Cardy, and E.~Tonni, {\it {Entanglement negativity in quantum
  field theory}},  {\em Phys.Rev.Lett.} {\bf 109} (2012) 130502,
  [\href{http://xxx.lanl.gov/abs/1206.3092}{{\tt arXiv:1206.3092}}].

\bibitem{Das:1997gg}
A.~K. Das, {\em {Finite temperature field theory}}.
\newblock World Scientific Publishing Company, 1997.
\newblock 404 p.

\bibitem{Headrick:2012fk}
M.~Headrick, A.~Lawrence, and M.~M. Roberts, {\it {Bose-Fermi duality and
  entanglement entropies}},  \href{http://xxx.lanl.gov/abs/1209.2428}{{\tt
  arXiv:1209.2428}}.

\bibitem{Casini:2005rm}
H.~Casini, C.~Fosco, and M.~Huerta, {\it {Entanglement and alpha entropies for
  a massive Dirac field in two dimensions}},  {\em J.Stat.Mech.} {\bf 0507}
  (2005) P07007, [\href{http://xxx.lanl.gov/abs/cond-mat/0505563}{{\tt
  cond-mat/0505563}}].

\bibitem{Aharony:1999ti}
O.~Aharony, S.~S. Gubser, J.~M. Maldacena, H.~Ooguri, and Y.~Oz, {\it {Large N
  field theories, string theory and gravity}},  {\em Phys.Rept.} {\bf 323}
  (2000) 183--386, [\href{http://xxx.lanl.gov/abs/hep-th/9905111}{{\tt
  hep-th/9905111}}].

\bibitem{Nishioka:2009un}
T.~Nishioka, S.~Ryu, and T.~Takayanagi, {\it {Holographic Entanglement Entropy:
  An Overview}},  {\em J. Phys.} {\bf A42} (2009) 504008,
  [\href{http://xxx.lanl.gov/abs/0905.0932}{{\tt arXiv:0905.0932}}].

\bibitem{Hayden:2011ag}
P.~Hayden, M.~Headrick, and A.~Maloney, {\it {Holographic Mutual Information is
  Monogamous}},  \href{http://xxx.lanl.gov/abs/1107.2940}{{\tt
  arXiv:1107.2940}}. 21 pages, 1 figure.

\bibitem{Israel:1976ur}
W.~Israel, {\it {Thermo field dynamics of black holes}},  {\em Phys.Lett.} {\bf
  A57} (1976) 107--110.

\bibitem{Hawking:1983aa}
S.~W. Hawking and D.~N. Page, {\it Thermodynamics of black holes in anti-de
  sitter space},  {\em Communications in Mathematical Physics} {\bf 87} (1983)
  577--588.

\bibitem{Kraus:2002iv}
P.~Kraus, H.~Ooguri, and S.~Shenker, {\it Inside the horizon with ads/cft},
  {\em Phys. Rev. D} {\bf 67} (Jun, 2003) 124022,
  [\href{http://xxx.lanl.gov/abs/hep-th/0212277}{{\tt hep-th/0212277}}].

\bibitem{Fidkowski:2004aa}
L.~Fidkowski, V.~Hubeny, M.~Kleban, and S.~Shenker, {\it The black hole
  singularity in ads/cft},  {\em Journal of High Energy Physics} {\bf 2004}
  (2004), no.~02 014, [\href{http://xxx.lanl.gov/abs/hep-th/0306170}{{\tt
  hep-th/0306170}}].

\bibitem{Aminneborg:1997pz}
S.~Aminneborg, I.~Bengtsson, D.~Brill, S.~Holst, and P.~Peldan, {\it {Black
  holes and wormholes in (2+1)-dimensions}},  {\em Class.Quant.Grav.} {\bf 15}
  (1998) 627--644, [\href{http://xxx.lanl.gov/abs/gr-qc/9707036}{{\tt
  gr-qc/9707036}}].

\bibitem{Brill:1998pr}
D.~Brill, {\it {Black holes and wormholes in (2+1)-dimensions}},
  \href{http://xxx.lanl.gov/abs/gr-qc/9904083}{{\tt gr-qc/9904083}}.

\bibitem{Krasnov:2000zq}
K.~Krasnov, {\it Holography and riemann surfaces},  {\em Adv.Theor.Math.Phys.}
  {\bf 4} (2000) 929--979, [\href{http://xxx.lanl.gov/abs/hep-th/0005106}{{\tt
  hep-th/0005106}}].

\bibitem{Krasnov:2003aa}
K.~Krasnov, {\it Black-hole thermodynamics and riemann surfaces},  {\em
  Classical and Quantum Gravity} {\bf 20} (2003), no.~11 2235,
  [\href{http://xxx.lanl.gov/abs/gr-qc/0302073}{{\tt gr-qc/0302073}}].

\bibitem{Skenderis:2009ju}
K.~Skenderis and B.~C. van Rees, {\it Holography and wormholes in 2+1
  dimensions},  {\em Commun.Math.Phys.} {\bf 301} (2011) 583--626,
  [\href{http://xxx.lanl.gov/abs/0912.2090}{{\tt arXiv:0912.2090}}].

\bibitem{Liu:2012eea}
H.~Liu and M.~Mezei, {\it {A Refinement of entanglement entropy and the number
  of degrees of freedom}},  \href{http://xxx.lanl.gov/abs/1202.2070}{{\tt
  arXiv:1202.2070}}.

\bibitem{Haag:1967aa}
R.~Haag, N.~M. Hugenholtz, and M.~Winnink, {\it {On the equilibrium states in
  quantum statistical mechanics}},  {\em Commun. Math. Phys.} {\bf 5} (1967),
  no.~3 215--236.

\end{thebibliography}\endgroup

\end{document}